\documentclass[aps,prd,preprint,groupedaddress,nofootinbib]{revtex4-1}

\usepackage{graphicx}
\usepackage{epsfig}
\usepackage{bm}
\usepackage{amsfonts}
\usepackage{amsmath}
\usepackage{amssymb}
\usepackage{multirow}

\begin{document}



\title{Warm DBI inflation with constant sound speed}





\author{S. Rasouli$^{1}$, K. Rezazadeh$^{1}$, A. Abdolmaleki$^{2}$, and K. Karami$^{1}$}

\address{
$^{1}$Department of Physics, University of Kurdistan, Pasdaran Street, P.O. Box 66177-15175, Sanandaj, Iran\\
$^{2}$Research Institute for Astronomy and Astrophysics of Maragha (RIAAM), P.O. Box 55134-441, Maragha, Iran
}


\date{\today}


\begin{abstract}

We study inflation with the Dirac-Born-Infeld (DBI) noncanonical scalar field in both the cold and warm scenarios. We consider the Anti-de Sitter warp factor $f(\phi)=f_{0}/\phi^{4}$ for the DBI inflation and check viability of the quartic potential $V(\phi)=\lambda\phi^{4}/4$ in light of the Planck 2015 observational results. In the cold DBI setting, we find that the prediction of this potential in the $r-n_s$ plane is in conflict with Planck 2015 TT,TE,EE+lowP data. This motivates us to focus on the warm DBI inflation with constant sound speed. We conclude that in contrary to the case of cold scenario, the $r-n_s$ result of warm DBI model can be compatible with the 68\% CL constraints of Planck 2015 TT,TE,EE+lowP data in the intermediate and high dissipation regimes, whereas it fails to be observationally viable in the weak dissipation regime. Also, the prediction of this model for the running of the scalar spectral index $dn_s/d\ln k$ is in good agreement with the constraint of Planck 2015 TT,TE,EE+lowP data. Finally, we show that the warm DBI inflation can provide a reasonable solution to the swampland conjecture that challenges the de Sitter limit in the standard inflation.

\end{abstract}


\maketitle




\section{Introduction}
\label{section:introduction}

Consideration of a short period of rapid inflationary expansion before the radiation dominated era can provide reasonable explanation for the well-known puzzles of the Hot Big Bang cosmology \cite{Starobinsky1980, Sato1981, Sato1981-2, Guth1981, Linde1982, Albrecht1982, Linde1983}. Also, the quantum fluctuations during inflation can lead to the perturbations whose we can see the imprints on the large-scale structure (LSS) formation and the anisotropies of cosmic microwave background (CMB) radiation \cite{Mukhanov1981, Hawking1982, Starobinsky1982, Guth1982}. Inflationary models generally predict an almost scale-invariant power spectrum for the primordial perturbations. This prediction has been confirmed by the recent observational data from the Planck satellite \cite{Planck2015, Planck2015non-Gaussianity}. Although, these observational data as well support the inflation paradigm, so far we cannot determine the dynamics of inflation exclusively.

In the standard inflationary scenario, a canonical scalar field minimally coupled to the Einstein gravity, is employed to explain the accelerated expansion of the universe during inflation. The scalar field responsible for inflation is called ``inflaton''. Unfortunately, we don't have a suitable candidate to play the role of inflaton in the standard canonical inflationary setting. After the discovery of the Higgs boson \cite{Higgs1964} at the Large Hadron Collider (LHC) at CERN \cite{ATLAS2012, CMS2012}, it is conceivable to consider it as an inflaton candidate. The scalar potential of the Higgs boson in the standard model of particle physics behaves asymptotically like the self-interacting quartic potential $V(\phi)=\lambda\phi^{4}/4$ in renormalizable gauge field theories \cite{Pich2007, Patrignani2016}. But unfortunately, this potential suffers from some critical problems in the standard inflationary setting. First, it leads to large values for the tensor-to-scalar ratio which are in conflict with the current observational bounds imposed by the Planck 2015 data \cite{Planck2015}. Second, the value of the dimensionless constant $\lambda$ deduced from CMB normalization is of order $\sim 10^{-13}$, which is anomalously far from the Higgs coupling $\lambda\simeq0.13$ coming from the experimental searches \cite{ATLAS2012, CMS2012}. Also, taking into account the experimental bound $\lambda\simeq0.13$, leads to rather large values for the second slow-roll parameter $\eta$, which disrupts the slow-roll conditions in the standard inflationary scenario \cite{Germani2010}. This is generally so-called the $\eta$-problem in the standard inflation \cite{Copeland1994}. However, so far some theoretical attempts have been done to resolve the problems of the Higgs boson to be regarded as the inflaton \cite{Germani2010, Cervantes-Cota1995, Bezrukov2008, Germani2017, Calmet2017, Ballesteros2017}.

In addition to the problems mentioned above, there is another subject proposed recently that challenges the consistency of the standard scenario of inflation. This problem is the swampland conjecture \cite{Agrawal2018, Ooguri2018} and it results from the string theory considerations in the early universe. The swampland conjecture implies that the validity of the de Sitter (dS) limit is in contrast with the slow-roll conditions in the standard inflation.

The celebrity of the string theory as a fundamental model for physical phenomena motivates us to search for an inflaton candidate in the prospect of this theory. An interesting inflaton candidate inspired from the string theory is suggested in the context of Dirac-Born-Infeld (DBI) inflation \cite{Silverstein2004, Alishahiha2004}. This setup proposes that the role of inflaton can be accomplished by the radial coordinate of a D3-brane moving in a warped region (throat) of a compactification space. The brane behaves like a point-like object and according to the direction of its motion in the warped space, there are two versions of brane inflation, namely the ``ultraviolet'' (UV) model \cite{Silverstein2004, Alishahiha2004}, and the ``infrared'' (IR) model \cite{Chen2005, Chen2005-2}. In the UV model, the inflaton moves from the UV side of the warped space to the IR side, while in the IR model, it moves in the inverse direction. Additionally, there exist a speed limit on the inflaton motion in the warped space. The speed limit is affected by the brane speed and the warp factor of the throat. Because of the speed limit, a parameter $\gamma$ is introduced and it is analogous to the Lorentz factor in special relativity. When the speed of brane approaches the speed limit, the parameter $\gamma$ can increase to arbitrarily large values ($\gamma \gg 1$), that we call this case as the ``relativistic'' regime. On the other side, in the ``non-relativistic'' regime, the brane speed is much less than the speed limit, giving $\gamma \to 1$.

The DBI inflation can be included in the class of $k$-inflation models in which inflation is driven by a noncanonical scalar field \cite{Armendariz-Picon1999, Garriga1999, Mukhanov2006, Helmer2006, Taveras2008, Bose2009, Bose2009-2, Franche2010, Franche2010-2, Devi2011, Unnikrishnan2012, Rezazadeh2015, Rezazadeh2017-2}. An outstanding feature of $k$-inflation is that in this model, the sound speed $c_s$ of the scalar perturbations can be less than the light speed. As a result, $k$-inflation models are capable to provide small tensor-to-scalar ratio favored by the latest observational data from the Planck 2015 collaboration \cite{Planck2015}. In addition, this feature can provide large non-Gaussianities \cite{Alishahiha2004, Chen2007, Li2008, Tolley2010, Seery2005, Panotopoulos2007} which discriminates between $k$-inflation and the standard canonical inflation that predicts undetectably small non-Gaussianities \cite{Maldacena2003, Acquaviva2003, Rigopoulos2005, Creminelli2004}. In the DBI inflation, the sound speed is equal to inverse of the Lorentz factor, $c_{s}=1/\gamma$. Thus, in the relativistic and non-relativistic regimes, we have $c_{s}\ll1$ and $c_{s}\sim1$, respectively.

In the standard inflationary scenario, the interaction of inflaton field with other fields is neglected during inflation. So, the universe remains cold during inflation, and an additional process called ``reheating'' should be introduced to the final stages of inflation to make possible for the universe, so that it can transit from the accelerating phase of the inflationary era to the decelerating phase of the radiation epoch \cite{Basset2006}. The details of the reheating process are unknown to us so far.

Alternatively, in the warm inflation scenario, the interaction between the inflaton and other fields has a dynamical role during inflation \cite{Berera1995, Berera1995-2}. As a result, the energy density of the inflaton field can be transformed to the energy density of the radiation field, so that inflation can terminate without resorting to any additional reheating process \cite{Berera1997}. Also, note that in the cold and warm inflations, respectively, the quantum fluctuations of the inflaton field \cite{Mukhanov1981, Hawking1982, Starobinsky1982, Guth1982} and thermal fluctuations of the radiation field \cite{Berera2000, Taylor2000} are responsible for the anisotropies of the CMB radiation as well as the LSS formation.

Dissipative effects of warm inflation modifies the scalar primordial power spectrum in various ways. The first modification appears in the equation of the inflaton fluctuations so that in the warm inflation scenario, it cast into a Langevin equation governing the dynamics of the perturbed inflaton field \cite{Berera1995, Berera1995-2, Berera2000, Taylor2000, Hall2004, Moss2007, Graham2009, Ramos2013}. The second modification stems from the fact that in enough high temperatures, the distribution of the inflaton particles may deviate from the vacuum phase space distribution and tends toward the excited Bose-Einstein distribution \cite{Bartrum2014, Bastero-Gil2014}. Finally, the last one is related to the cases in which duo to the temperature dependence of the dissipation coefficient, the inflaton and radiation fluctuations are coupled to each other, and this modification specially becomes important in the high dissipative regime \cite{Bastero-Gil2016, Benetti2017}.

The required condition for realization of the warm inflation scenario is that the temperature $T$ of the universe should be larger than the Hubble expansion rate $H$ ($T>H$). This is necessary in order for the dissipation to potentially affect both the inflaton background dynamics, and the primordial power spectrum of the field fluctuations \cite{Bastero-Gil2009}. A few years after the original proposal for warm inflation, it was realized that the condition $T>H$ cannot be easily provided in conventional models \cite{Berera1998, Yokoyama1999}. Indeed, the inflaton could not couple directly with light fields easily. Moreover, a direct coupling to light fields can give rise to significant thermal corrections to the inflaton mass, so that the slow-roll regime is disturbed if $T>H$. However, it was shown in the next promotions that the indirect coupling of inflaton to light degree of freedoms can provide successful models of warm inflation. The required conditions for these class of warm inflation models can be realized in special scenarios such as the case of the brane models \cite{Bastero-Gil2011}. More recently, a new mechanism for warm inflation has been proposed in \cite{Bastero-Gil2016}, where warm inflation can be driven by an inflaton field coupled directly to a few light fields. In this scenario, the role of inflaton is played by a ``Little Higgs'' boson which is a pseudo Nambu-Goldstone boson (PNGB) of a broken gauge symmetry \cite{Arkani-Hamed2001, Schmaltz2005, Kaplan2004, Arkani-Hamed2003}.

In this paper, within the framework of cold and warm DBI inflation we consider the Anti-de Sitter (AdS) warp factor $f(\phi)=f_{0}/\phi^{4}$, and check viability of the quartic potential $V(\phi)=\lambda\phi^{4}/4$ in light of the Planck 2015 data \cite{Planck2015}. First, in Secs. \ref{section:cold} and \ref{section:cold,quartic_potential}, we present the background equations valid in the slow-roll approximation in the framework of cold DBI scenario, and examine inflation with the quartic potential $V(\phi)=\lambda \phi^4/4$. We show that in the cold DBI inflation, like the standard scenario, the result of quartic potential cannot be compatible with the Planck 2015 constraints \cite{Planck2015}. Then, in Secs. \ref{section:warm} and \ref{section:warm,quartic_potential}, we investigate the warm DBI inflation and check the viability of the quartic potential to see whether this potential can be resurrected in light of the Planck 2015 results. Section \ref{dissipation parameter} is devoted to some regards on the dissipation parameter in our model. In Sec. \ref{section:swampland}, we examine the possibility of resolving the swampland conjecture problem in our model. Finally, in Sec. \ref{section:conclusions} we summarize our concluding remarks.


\section{Cold DBI inflation}
\label{section:cold}

In this section, we first present the basic equations in the cold DBI inflationary setting in the slow-roll approximation. The dynamics of a D3-brane in the warped space is determined by the DBI noncanonical action
\begin{equation}
 \label{S}
 S=\int d^{4}x\sqrt{-g}\left[\frac{M_{P}^{2}}{2}R+2\mathcal{L}(X,\phi)\right],
\end{equation}
where $M_{P}=1/\sqrt{8\pi G}$ is the reduced Planck mass, and $R$ is the Ricci curvature scalar. Also, $\mathcal{L}(X,\phi)$ is the DBI Lagrangian which is a function of the inflaton scalar field $\phi$ and the canonical kinetic term $X=-\partial^{\mu}\phi\partial_{\mu}\phi/2$. The DBI Lagrangian is in the form
\begin{equation}
 \label{{mathcal}{L}}
 \mathcal{L}(X,\phi)=f^{-1}(\phi)\left[1-\sqrt{1-2f(\phi)X}\right]-V(\phi),
\end{equation}
where $f(\phi)$ and $V(\phi)$ are the warp factor and the potential energy of the inflaton, respectively. We consider a homogeneous and isotropic universe with the spatially flat FRW metric, and assume that the universe is filled by a perfect fluid with the energy-stress tensor $T^{\mu}_{~~\nu}=\mathrm{diag}\left(-\rho_{\phi},p_{\phi},p_{\phi},p_{\phi}\right)$. As a result, the canonical kinetic term is transformed to $X=\dot{\phi}^{2}/2$. The energy density and pressure of the DBI scalar field are respectively given by
\begin{align}
 \label{{rho}_{phi}}
 \rho_{\phi} &\equiv2X\mathcal{L}_{,{\rm X}}-\mathcal{L}=\frac{\gamma-1}{f(\phi)}+V(\phi),
 \\
 \label{p_{phi}}
 p_{\phi} &\equiv\mathcal{L}=\frac{\gamma-1}{\gamma f(\phi)}-V(\phi),
\end{align}
where $\mathcal{L}_{,{\rm X}}\equiv\partial\mathcal{L}/\partial X$, and $\gamma$ is the Lorentz factor defined as
\begin{equation}
 \label{{gamma}}
 \gamma\equiv\frac{1}{\sqrt{1-f(\phi)\dot{\phi}^{2}}}.
\end{equation}
In the relativistic regime, the speed of the brane motion, $|\dot{\phi}|$, approaches to the speed limit $f^{-1/2}(\phi)$, and hence the Lorentz factor grows without bound, $\gamma\gg1$. ‌Instead, in the non-relativistic regime, the speed of the brane motion is much less than the speed limit, $|\dot{\phi}|\ll f^{-1/2}(\phi)$, and consequently the Lorentz factor approaches unity, $\gamma \simeq 1$.

For the DBI scalar field model, the sound speed is given by
\begin{equation}
 \label{c_s}
 c_{s}^{2}\equiv\frac{\partial p_{\phi}/\partial X}{\partial\rho_{\phi}/\partial X}=\frac{1}{\gamma^{2}}=1-f(\phi)\dot{\phi}^{2},
\end{equation}
which specifies the propagation speed of the density perturbations among the background, and it should be real and subluminal, $0<c_{s}^{2}\leq1$ \cite{Franche2010}.

In the cold DBI inflationary setting, the first Friedmann equation is
\begin{equation}
 \label{H,{rho}_{phi}}
 H^{2}=\frac{1}{3M_{P}^{2}}\rho_{\phi},
\end{equation}
where $H\equiv\dot{a}/a$ is the Hubble parameter specifying the expansion rate of the scale factor $a$ of the universe. Here, the dot denotes the derivative with respect to the cosmic time $t$. In the cold inflation scenario, it is presumed that the inflaton field does not interact with other fields during inflationary period, and thus creation of particles is not possible during inflation. As a result, the energy conservation law implies
\begin{equation}
 \label{{dot}{{rho}}_{phi}}
 \dot{\rho}_{\phi}+3H\left(\rho_{\phi}+p_{\phi}\right)=0.
\end{equation}
Inserting $\rho_{\phi}$ from Eq. (\ref{{rho}_{phi}}) into the above equation, we obtain the equation of motion of the inflaton as
\begin{equation}
 \label{{ddot}{{phi}}}
 \ddot{\phi}+\frac{3H}{\gamma^{2}}~\dot{\phi}+\frac{V'(\phi)}{\gamma^{3}}
 +\frac{f'(\phi)(\gamma+2)(\gamma-1)}{2f(\phi)\gamma(\gamma+1)}~\dot{\phi}^{2}=0,
\end{equation}
which a prime denotes the derivative with respect to the scalar field $\phi$. Note that for $\gamma=1$ Eq. (\ref{{ddot}{{phi}}}) reduces to the well known relation in the standard inflation.

Combining Eqs. (\ref{{rho}_{phi}}), (\ref{p_{phi}}), (\ref{{gamma}}), (\ref{H,{rho}_{phi}}), and (\ref{{dot}{{rho}}_{phi}}), we reach
\begin{equation}
 \label{{dot}{H}}
 \dot{H}=-\frac{\gamma}{2M_{P}^{2}}\dot{\phi}^{2}.
\end{equation}
This equation is exact, and it is very useful in the study of cold DBI inflation.

Now, following \cite{Peiris2007, Spalinski2007-3, Bessada2009}, we introduce the slow-roll parameters as
\begin{align}
 \label{{epsilon}}
 \epsilon &\equiv-\frac{\dot{H}}{H^{2}},
 \\
 \label{{eta}}
 \eta &\equiv\frac{\dot{\epsilon}}{H\epsilon},
 \\
 \label{{kappa}}
 \kappa &\equiv\frac{\dot{c}_{s}}{Hc_{s}}.
\end{align}
The parameters $\epsilon$ and $\eta$ are called the first and second Hubble slow-roll parameters, respectively, and quantify the variation of the Hubble parameter and its temporal derivative in duration of inflation. The parameter $\kappa$ is so-called the sound slow-roll parameter, because it describes the evolution rate of the sound speed. These parameters are much less than unity, $\epsilon,\,\eta,\,\kappa\ll1$, in the regime of slow-roll inflation.

It is convenient to apply the Hamilton-Jacobi formalism \cite{Muslimov1990, Salopek1990, Kinney1997}, in which the Hubble parameter $H$ is taken as a function of the inflaton field $\phi$. The Hamilton-Jacobi formalism in the context of DBI inflation, at first was introduced in \cite{Silverstein2004}, and then it was discussed in some details in \cite{Spalinski2007}. Applying this formalism, we can also define the Hamilton-Jacobi slow-roll parameters \cite{Peiris2007, Bessada2009} as
\begin{align}
 \label{{tilde}{{epsilon1}}}
 \tilde{\epsilon} &\equiv\frac{2M_{P}^{2}}{\gamma}\left(\frac{H'}{H}\right)^{2},
 \\
 \label{{tilde}{{eta}}}
 \tilde{\eta} &\equiv\frac{2M_{P}^{2}}{\gamma}\frac{H''}{H},
 \\
 \label{{tilde}{{kappa}}}
 \tilde{\kappa} &\equiv\frac{2M_{P}^{2}}{\gamma}\frac{H'}{H}\frac{\gamma'}{\gamma},
\end{align}
where the prime denotes the derivative with respect to $\phi$. In the cold DBI setting, using Eqs. (\ref{{gamma}}), (\ref{c_s}), and (\ref{{dot}{H}}) one can show that the Hubble slow-roll parameters and the Hamilton-Jacobi ones are related together as \cite{Peiris2007, Bessada2009}
\begin{align}
 \label{{epsilon},{tilde}{{epsilon}}}
 \epsilon &=\tilde{\epsilon},
 \\
 \label{{eta},{tilde}{{eta}}}
 \eta &=2\tilde{\epsilon}-2\tilde{\eta}-\tilde{\kappa},
 \\
 \label{{kappa},{tilde}{{kappa}}}
 \kappa &=\tilde{\kappa}.
\end{align}

Using Eqs. (\ref{{rho}_{phi}}), (\ref{c_s}), (\ref{{dot}{H}}), and (\ref{{epsilon}}) one can obtain
\begin{equation}
 \label{H,V,{epsilon}}
 1=\frac{2}{3}\frac{\epsilon}{1+c_{s}}+\frac{V(\phi)}{3M_{P}^{2}H^{2}},
\end{equation}
which in the slow-roll regime, the slow-roll parameter $\epsilon$ is much less than unity, and hence the first term on the right hand side of the above equation can be neglected versus the second one. Therefore, in the DBI cold inflation, the first Friedmann equation in the slow-roll approximation yields \cite{Chen2007, Miranda2012, Amani2017}
\begin{equation}
 \label{H,V}
 H^{2}\approx\frac{1}{3M_{P}^{2}}V(\phi),
\end{equation}
which is same as that obtained in the standard canonical inflation. Also, in the slow-roll approximation, the first term on the left hand side of Eq. (\ref{{ddot}{{phi}}})
can be ignored in comparison with the other terms, and consequently using Eqs. (\ref{{gamma}}) and (\ref{c_s}), we will have \cite{Miranda2012, Amani2017}
\begin{equation}
 \label{{dot}{{phi}}}
 3H\dot{\phi}+c_{s}V'(\phi) \approx0.
\end{equation}
In should be noted that in the case $c_s=1$, the above equation reduces to the one being valid in the standard canonical inflation in the slow-roll approximation.

The size of the universe during inflation is usually expressed in terms of the $e$-fold number
\begin{equation}
 \label{N}
 N\equiv\ln\left(\frac{a_{e}}{a}\right),
\end{equation}
where $a_{e}$ is the scale factor of the universe at the end of inflation. It is believed that the largest scales in the observable universe left the horizon at $N_{*}\approx50-60$ $e$-folds from the end of inflation \cite{Liddle2003, Dodelson2003}. Using Eqs. (\ref{{dot}{H}}) and (\ref{N}) one can obtain the following relation for the number of $e$-foldings
\begin{equation}
 \label{dN}
 dN=-Hdt=-\frac{H}{\dot{\phi}}~d\phi=-\frac{1}{M_{P}}\sqrt{\frac{\gamma}{2\epsilon}}\,\left|d\phi\right|.
\end{equation}
To determine $\left|d\phi\right|$, it is required to specify whether the inflaton field increases or decreases during inflation, and obviously this depends on the shape of inflationary potential. In other words, in the models for which $V'(\phi)<0$ and $V'(\phi)>0$, one should take $\left|d\phi\right|=d\phi$ and $\left|d\phi\right|=-d\phi$, respectively.

In the following, we present the basic equations governing the primordial perturbations in the setup of cold DBI inflation. It is believed that two types of perturbations are generated during inflation, namely the scalar and tensor perturbations. The scalar perturbations are the origin of the density fluctuations leading to the LSS formation and the CMB temperature anisotropies \cite{Mukhanov1981, Hawking1982, Starobinsky1982, Guth1982}. The tensor perturbations are the origin of the primordial gravitational waves which affect the polarization anisotropies of the  CMB radiation. The power spectrum of the scalar perturbations in the framework of cold DBI inflation is given by \cite{Peiris2007, Bessada2009}
\begin{equation}
 \label{{mathcal}{P}_s}
 \mathcal{P}_{s}=\left.\frac{H^{2}}{8\pi^{2}M_{P}^{2}c_{s}\epsilon}\right|_{c_{s}k=aH}.
\end{equation}
The above expression should be evaluated at the sound horizon exit for which $c_{s}k=aH$, where $k$ denotes the comoving wavenumber. To quantify the scale-dependence of the scalar power spectrum, we define the scalar spectral index as
\begin{equation}
 \label{n_s,{definition}}
 n_{s}-1\equiv\frac{d\ln\mathcal{P}_{s}}{d\ln k}.
\end{equation}
Since in the slow-roll inflation, the Hubble parameter $H$ and the sound speed $c_s$ vary significantly slower than the scale factor $a$ \cite{Garriga1999}, so the relation $c_{s}k=aH$ yields
\begin{equation}
 \label{d{ln}k}
 d\ln k\approx Hdt=-dN.
\end{equation}
With the help of Eqs. (\ref{{epsilon}}), (\ref{{eta}}), (\ref{{kappa}}), (\ref{{epsilon},{tilde}{{epsilon}}}), (\ref{{eta},{tilde}{{eta}}}), (\ref{{kappa},{tilde}{{kappa}}}), (\ref{{mathcal}{P}_s}), and (\ref{d{ln}k}), we can obtain the scalar spectral index for the cold DBI inflation as \cite{Peiris2007, Bessada2009}
\begin{equation}
 \label{n_s}
 n_{s}=1-4\tilde{\epsilon}+2\tilde{\eta}-2\tilde{\kappa}.
\end{equation}
The tensor power spectrum in the DBI model is the same as that of the standard canonical inflation, because both of these inflationary models are based on the Einstein gravity, and it is given by \cite{Peiris2007, Bessada2009}
\begin{equation}
 \label{{mathcal}{P}_t}
 \mathcal{P}_{t}=\left.\frac{2H^{2}}{\pi^{2}M_{P}^{2}}\right|_{k=aH}.
\end{equation}
In contrast with the scalar power spectrum which should be evaluated at the sound horizon crossing ($c_{s}k=aH$), the tensor power spectrum should be calculated at the usual Hubble horizon exit determined by $k=aH$. Nevertheless, the difference between the sound and Hubble horizon exit times can be ignored in the slow-roll approximation \cite{Garriga1999, Unnikrishnan2012}. The scale-dependence of the tensor power spectrum is determined by the tensor spectral index
\begin{equation}
 \label{n_t,{definition}}
 n_{t}\equiv\frac{d\ln\mathcal{P}_{t}}{d\ln k}.
\end{equation}
In analogy with the approach followed to derive $n_s$ in Eq. (\ref{n_s}), we can obtain the tensor spectral index for the cold DBI inflationary model as \cite{Peiris2007, Bessada2009}
\begin{equation}
 \label{n_t}
 n_{t}=-2\epsilon.
\end{equation}
An important inflationary observable is the tensor-to-scalar ratio defined as
\begin{equation}
 \label{r,{definition}}
 r\equiv\frac{\mathcal{P}_{t}}{\mathcal{P}_{s}}.
\end{equation}
This quantity has a special significance in discriminating between different inflationary models. In fact, combining of the observational bound on $r$ and $n_s$ provides a powerful criterion to distinguish between viable inflationary models in light of the observational data. Inserting Eqs. (\ref{{mathcal}{P}_s}) and (\ref{{mathcal}{P}_t}) into  (\ref{r,{definition}}), we get
\begin{equation}
 \label{r}
 r=16c_{s}\epsilon.
\end{equation}

Another important discriminator which has a wide usage in checking viability of inflationary models, is the non-Gaussianity parameter. In the context of DBI inflation, the primordial non-Gaussianity parameter is given by \cite{Alishahiha2004, Chen2007, Li2008, Tolley2010}
\begin{equation}
\label{f_{NL}^{DBI}}
f_{{\rm NL}}^{{\rm DBI}}=-\frac{35}{108}\left(\frac{1}{c_{s}^{2}}-1\right).
\end{equation}
The observational constraint from Planck 2015 data on the equilateral non-Gaussianity parameter is $f_{{\rm NL}}^{{\rm DBI}}=15.6\pm37.3$ (68\% CL, Planck 2015 T+E) \cite{Planck2015non-Gaussianity}. The Planck 2015 collaboration have applied their 95\% CL constraints from T+E data on Eq. (\ref{f_{NL}^{DBI}}), and find the lower bound $c_{s}\geq0.087$ on the sound speed of the DBI inflation \cite{Planck2015non-Gaussianity}.


\section{Cold DBI inflation with the quartic potential}
\label{section:cold,quartic_potential}

In the previous section, we presented the necessary relations governing the inflationary observables in the context of cold DBI inflation. Here, we use these results to examine the self-interacting quartic potential
\begin{equation}
 \label{V,{quartic}}
 V(\phi)\text{=}\frac{1}{4}\lambda\phi^{4},
\end{equation}
where $\lambda$ is a dimensionless parameter. This potential can drive a simple chaotic inflation \cite{Linde1983}, and it has a substantial importance for the theoretical standpoint, because it emerges in almost all renormalizable gauge field theories \cite{Pich2007}, in which the gauge is spontaneously broken through the Higgs mechanism \cite{Higgs1964}. This potential further possesses some appealing reheating properties \cite{Liddle2003, Ford1987}, which makes it very important in the inflationary landscape. But unfortunately, in the standard inflation, the prediction of this potential in $r-n_s$ plane lies completely outside the region favored by the recent Planck data \cite{Planck2015}. Here, we are interested to examine the compatibility of this potential with the observational data within the setting of cold DBI inflation.

In our work, we consider the AdS warp factor \cite{Silverstein2004, Alishahiha2004, Aharony2000}
\begin{equation}
 \label{f}
 f(\phi)=\frac{f_{0}}{\phi^{4}},
\end{equation}
where $f_{0}$ is a positive dimensionless constant. Considering the quartic potential (\ref{V,{quartic}}), the Friedmann equation in the slow-roll approximation, Eq. (\ref{H,V}), gives
\begin{equation}
 \label{H,{quartic}}
 H=\frac{\sqrt{\lambda}}{2\sqrt{3}M_{P}}\,\phi^{2}.
\end{equation}
The scalar spectral index and the tensor-to-scalar ratio follow from Eqs. (\ref{n_s}) and (\ref{r}), respectively, as
\begin{align}
 \label{n_s,{quartic}}
 n_{s} &= 1-\frac{8\sqrt{3}M_{P}^{2}\big(9\phi^{2}+8M_{P}^{2}\tilde{f}_{0}\big)}{\phi\big(3\phi^{2}+4M_{P}^{2}\tilde{f}_{0}\big)^{3/2}},
 \\
 \label{r,{quartic}}
 r &= \frac{384M_{P}^{2}}{3\phi^{2}+4M_{P}^{2}\tilde{f}_{0}},
\end{align}
where we have defined $\tilde{f}_{0}\equiv f_{0}\lambda$.

From Eqs. (\ref{{dot}{{phi}}}) and (\ref{dN}), we arrive at
\begin{equation}
 \label{dN/d{phi},{quartic}}
 \frac{dN}{d\phi}=\frac{1}{4M_{P}^{2}}\sqrt{\phi^{2}+\frac{4}{3}M_{P}^{2}\tilde{f}_{0}}~.
\end{equation}
Taking into account the initial condition $N_{e}=N\left(\phi_{e}\right)=0$ at the end of inflation, the solution of the above differential equation is
\begin{equation}
 \label{N,{phi},{quartic}}
 N=\frac{1}{24M_{P}^{2}}\left.\left[\phi\sqrt{9\phi^{2}+12M_{P}^{2}\tilde{f}_{0}}+4M_{P}^{2}\tilde{f}_{0}\ln\left(\sqrt{9\phi^{2}+12M_{P}^{2}\tilde{f}_{0}}+3\phi\right)\right]\right|_{\phi_{e}}^{\phi}.
\end{equation}
The first slow-roll parameter (\ref{{tilde}{{epsilon1}}}) for the quartic potential (\ref{V,{quartic}}) turns into
\begin{equation}
 \label{{tilde}{{epsilon}}}
 \tilde{\epsilon}=\frac{8M_{P}^{2}}{\phi\sqrt{\phi^{2}+\frac{4}{3}M_{P}^{2}\tilde{f}_{0}}}.
\end{equation}
By putting $\tilde{\epsilon}=1$ at the end of inflation, the inflaton scalar field reads
\begin{equation}
 \label{{phi}_e,{quartic}}
 \phi_{e}=M_{P}\sqrt{\frac{2}{3}\left(\sqrt{\tilde{f}_{0}^{2}+144}-\tilde{f}_{0}\right)}.
\end{equation}

We use a numerical approach to solve Eq. (\ref{N,{phi},{quartic}}) and find $\phi$ in terms of $N$. Afterwards, we insert it in Eq. (\ref{n_s,{quartic}}) and (\ref{r,{quartic}}) to evaluate $n_s$ and $r$ at the epoch of horizon crossing with a given $N_*$. Consequently, the $r-n_s$ plot of this model is resulted in as shown by the dashed ($N_*=50$) and solid ($N_*=60$) green lines in Fig. \ref{figure:r,n_s}. Note that in the standard inflation, the scalar spectral index and tensor-to-scalar of the quartic potential (\ref{V,{quartic}}) are given by $n_{s}=1-3/N$ and $r=16/N$, respectively \cite{Okada2016}. Using these equations, the $r-n_s$ plot of this potential is shown by the blue line in Fig. \ref{figure:r,n_s}, that lies quite outside of the favored region of the Planck 2015 observational results \cite{Planck2015}. As it is apparent in Fig. \ref{figure:r,n_s}, the result of the quartic potential (\ref{V,{quartic}}) in the cold DBI inflation setting, like the standard canonical one, is completely inconsistent with the Planck 2015 observational results \cite{Planck2015}. In Fig. \ref{figure:r,n_s}, we have considered the parameter $\tilde{f}_{0}\geq 0$ as the varying parameter to plot the $r-n_s$ diagram of the DBI model. As the parameter $\tilde{f}_{0}$ decreases to zero, the sound speed $c_s$ approaches unity and consequently the $r-n_s$ diagram of the DBI inflation goes toward the result of the quartic potential (\ref{V,{quartic}}) in the standard framework.


\section{Warm DBI inflation with constant sound speed}
\label{section:warm}

In this section, we proceed to study the warm DBI inflation, and present the basic equations of this inflationary scenario. In warm inflation, the inflaton scalar field interacts with other fields, and dissipation plays a dynamical role during inflation. Because of the dissipative effects, the vacuum energy changes into the radiation energy, and therefore the conservation equations governing the energy densities of the inflaton $\rho_\phi$ and the radiation $\rho_{R}$ take the following forms
\begin{align}
 \label{{dot}{{rho}}_{phi},{warm}}
 \dot{\rho}_{\phi}+3H\left(\rho_{\phi}+p_{\phi}\right) &= -\Upsilon\left(\rho_{\phi}+p_{\phi}\right),
 \\
 \label{{dot}{{rho}}_R}
 \dot{\rho}_{R}+3H\left(\rho_{R}+p_{R}\right) &= \Upsilon\left(\rho_{\phi}+p_{\phi}\right),
\end{align}
where $\Upsilon$ is the dissipation parameter. The dissipation parameter is determined by the QFT, and in general it can be a function of the inflaton field and the temperature, or both of them \cite{Berera1998, Berera2001, Berera2005}. The first Friedmann equation in warm inflation reads
\begin{equation}
 \label{H,{rho}_{phi},{rho}_R}
 H^{2}=\frac{1}{3M_{P}^{2}}\left(\rho_{\phi}+\rho_{R}\right).
\end{equation}
We postulate that during warm inflation, the relativistic particles are thermalized, so that the radiation energy density satisfies the black body equation
\begin{equation}
 \label{{rho}_R,T}
 \rho_{R}=\alpha T^{4},
\end{equation}
where $T$ is the temperature of the thermalized bath, and $\alpha=\pi^{2}g_{*}/30$ is the Stefan-Boltzmann constant in which $g_*$ is the relativistic degrees of freedom, and in this paper we take it as $g_{*}\simeq228.75$ \cite{Bartrum2014}.

Note that using Eq. (\ref{c_s}), we can simply rewrite Eqs. (\ref{{rho}_{phi}}) and (\ref{p_{phi}}) as
\begin{align}
 \label{{rho}_{phi},c_s}
 \rho_{\phi} &=\frac{\dot{\phi}^{2}}{c_{s}\left(1+c_{s}\right)}+V(\phi),
 \\
 \label{p_{phi},c_s}
 p_{\phi} &=\frac{\dot{\phi}^{2}}{1+c_{s}}-V(\phi).
\end{align}
In the following, we consider the sound speed (\ref{c_s}) to be constant, i.e. $c_{s}=\mathrm{const.}$. If we take the derivative of Eq. (\ref{{rho}_{phi},c_s}), and replace the result into Eq. (\ref{{dot}{{rho}}_{phi},{warm}}), we will have
\begin{equation}
 \label{{ddot}{{phi}},{warm}}
 \frac{2\ddot{\phi}}{1+c_{s}}+3H\left(1+Q\right)\dot{\phi}+c_{s}V'\left(\phi\right)=0,
\end{equation}
where $Q$ is the dissipation ratio defined as
\begin{equation}
 \label{Q}
 Q\equiv\frac{\Upsilon}{3H}.
\end{equation}

It should be noted that the warm inflation occurs when the temperature of thermal bath dominates over the Hubble expansion rate, i.e. $T>H$ \cite{Bastero-Gil2009}. This requirement implies that during warm inflation, the thermal fluctuations of the radiation field dominates over the quantum fluctuations of the inflaton field, so that they can be regarded as the origin of the LSS formation and the CMB anisotropies \cite{Berera2000, Taylor2000}.

In our work to study the warm DBI inflation, we consider the slow-roll parameters as introduced in Eqs. (\ref{{tilde}{{epsilon1}}}), (\ref{{tilde}{{eta}}}), and (\ref{{tilde}{{kappa}}}) for the cold DBI inflation. In addition, following \cite{Cai2011}, we define the fourth slow-roll parameter as
\begin{equation}
 \label{{tile}{{sigma}}}
 \tilde{\sigma}\equiv2M_{P}^{2}c_{s}\frac{\Upsilon'}{\Upsilon}\frac{H'}{H},
\end{equation}
which quantifies the variation of the dissipation parameter $\Upsilon$. Also, here we consider the slow-roll regime of warm inflation in which the slow-roll parameters are very small relative to $1+Q$. However, around the final stages of warm inflation, usually the slow-roll conditions are violated, and the slow-roll parameters become comparable with $1+Q$. Furthermore, we note that at the beginning of inflation, the energy density of inflaton dominates over the radiation energy density, i.e. $\rho_{\phi} \gg\rho_{R}$, and inflation terminates when the energy density of the radiation field becomes comparable with the inflaton energy density. In the energy density of the scalar field it is supposed that during inflation the contribution of the potential energy dominates over the kinetic energy so that $\rho_\phi \approx V(\phi)$.  It is further assumed that production of the photons are quasi-stable in the sense that $\dot{\rho}_{R}\ll4H\rho_{R}$. Putting all of these assumptions together, we can show that Eqs. (\ref{{dot}{{rho}}_R}) , (\ref{H,{rho}_{phi},{rho}_R}), and (\ref{{ddot}{{phi}},{warm}}) reduce to
\begin{align}
 \label{H,{rho}_{phi},{warm}}
 & H^{2}\approx\frac{1}{3M_{P}^{2}} V(\phi),
 \\
 \label{{dot}{{phi}},{warm}}
 & 3H\left(1+Q\right)\dot{\phi}+c_{s}V'\left(\phi\right)\approx0,
 \\
 \label{{rho}_R,{dot}{{phi}}}
 & \rho_{R}\approx\frac{3Q\dot{\phi}^{2}}{4c_{s}}.
\end{align}
Now, we turn to study perturbations in the DBI warm inflation. In warm inflation, the origin of the scalar perturbations is the thermal fluctuation of the radiation field which appears as a thermal noise in the perturbed inflaton field equation. The evolution equation of the perturbed inflaton field coupled with the radiation fluctuations is described by a Langevin equation \cite{Gleiser1994, Berera2000, Taylor2000, Moss2007, Graham2009, Cai2011, Zhang2014, Zhang2015}. The effect of thermal noise decreases up to the time when the fluctuation amplitude freezes out, that commonly happens before the horizon exit \cite{Gleiser1994, Berera2000, Taylor2000, Moss2007, Graham2009, Cai2011, Zhang2014, Zhang2015}. In the DBI warm inflation, the curvature perturbations can be calculated beyond the sound horizon scales, and after matching to the perturbations frozen-out at the smaller thermal noise scales, the power spectrum of the curvature perturbations is obtained as \footnote{In this paper, we adopted the standard convention of \cite{Graham2009}, and therefore the expression of $\mathcal{P}_s$ in Eq. (\ref{{mathcal}{P}_s,{warm}}) is different to a factor $4\pi^2$ in the denominator relative to the one presented in \cite{Cai2011} which has followed the notation of \cite{Hall2004}.} \cite{Cai2011}
\begin{equation}
 \label{{mathcal}{P}_s,{warm}}
 \mathcal{P}_{s}=\frac{\sqrt{3}}{16\pi^{3/2}M_{P}^{2}}\frac{HT}{c_{s}\tilde{\epsilon}}(1+Q)^{5/2}.
\end{equation}
The tensor perturbations which are the source of the primordial gravitational waves, do not have any interaction with the thermal fluctuations. Therefore, the tensor perturbations are completely decoupled from the thermal perturbations, and consequently the tensor power spectrum in the warm inflation is the same as that of the cold inflation, which is given by Eq. (\ref{{mathcal}{P}_t}). Note that from Eqs. (\ref{{mathcal}{P}_s}) and (\ref{{mathcal}{P}_s,{warm}}) we obtain $\mathcal{P}_{s}^{\rm (warm)}/\mathcal{P}_{s}^{\rm (cold)}\propto \frac{T}{H}(1+Q)^{5/2}$ which is greater than one because $T/H>1$ and $Q>0$. This shows that the amplitude of the scalar perturbations in the warm inflation is greater than that of the cold inflation, while the tensor amplitude $\mathcal{P}_{t}$ remains unchanged. Consequently, in the warm inflation the tensor-to-scalar ratio $r=\mathcal{P}_{t}/\mathcal{P}_{s}$ becomes smaller in comparison with the cold scenario.

At the level of nonlinear perturbations, as it has been discussed in \cite{Cai2011}, for sufficiently small values of $c_s$, the contribution of the inflaton perturbations dominates over the thermal contribution in the cubic order Lagrangian, and hence the equilateral non-Gaussianity parameter is obtained as in the cold DBI inflation, which is given by Eq. (\ref{f_{NL}^{DBI}}). Therefore, again the Planck 2015 constraints on the primordial non-Gaussianities \cite{Planck2015non-Gaussianity} lead to the bound $c_{s}\geq0.087$ on the sound speed of our model.


\section{Warm DBI inflation with the quartic potential}
\label{section:warm,quartic_potential}

As we saw in Sec. \ref{section:cold,quartic_potential}, the quartic potential (\ref{V,{quartic}}) fails to be consistent with the Planck 2015 data in the framework of cold DBI inflation. Then, we are motivated to examine the consistency of this model in the warm DBI inflationary setting. In our investigation, we again adopt the AdS warp factor (\ref{f}), but now we make the further assumption that the sound speed $c_s$ to be constant during inflation. The idea of constant sound speed in study of the DBI inflation has been regarded in \cite{Spalinski2008, Tsujikawa2013, Amani2017}, and it has been shown that it leads to considerable simplification in calculations. We employ this assumption in our work, because the definite form of the dissipation parameter $\Upsilon$ has not been determined so far, and to specify it, we require further studies for the inflaton interactions in the thermal bath.

By the use of Eqs. (\ref{{tilde}{{epsilon1}}}), (\ref{{tilde}{{eta}}}), and (\ref{{tilde}{{kappa}}}) for the quartic potential (\ref{V,{quartic}}), the slow-roll parameters turn into
\begin{align}
 \label{{tilde}{{epsilon}},{quartic}}
 \tilde{\epsilon} &= \frac{8M_{P}^{2}c_{s}}{\phi^{2}},
 \\
 \label{{tilde}{{eta}},{quartic}}
 \tilde{\eta} &= \frac{4M_{P}^{2}c_{s}}{\phi^{2}},
 \\
 \label{{tilde}{{kappa}},{quartic}}
 \tilde{\kappa} &= 0.
\end{align}
Now, using Eq. (\ref{c_s}) for the AdS warp factor (\ref{f}), we arrive at
\begin{equation}
 \label{{dot}{{phi}},{quartic},{warm}}
 \dot{\phi}=-\sqrt{\frac{1-c_{s}^{2}}{f_{0}}}~\phi^{2}.
\end{equation}
Inserting this into Eqs. (\ref{H,{rho}_{phi},{warm}}), (\ref{{dot}{{phi}},{warm}}), and (\ref{{rho}_R,{dot}{{phi}}}), we respectively get
\begin{align}
 \label{H,{phi},{quartic},{warm}}
 H &= \frac{1}{2M_{P}}\sqrt{\frac{\tilde{f}_{0}c_{s}-4c_{s}+4}{3f_{0}c_{s}}}\,\phi^{2},
 \\
 \label{Q,{phi},{quartic},{warm}}
 Q &= \frac{2M_{P}\tilde{f}_{0}c_{s}^{3/2}}{\sqrt{3\left(1-c_{s}^{2}\right)\left[c_{s}\left(\tilde{f}_{0}-4\right)+4\right]}\,\phi}-1,
 \\
 \label{T,{phi},{quartic},{warm}}
 T &= \left[\frac{1-c_{s}^{2}}{4\alpha c_{s}f_{0}}\,\phi^{3}\left(\frac{2\sqrt{3}M_{P}\tilde{f}_{0}c_{s}^{3/2}}{\sqrt{\left(1-c_{s}^{2}\right)\left[c_{s}\left(\tilde{f}_{0}-4\right)+4\right]}}-3\phi\right)\right]^{1/4},
\end{align}
where $\tilde{f}_{0}\equiv f_{0}\lambda$. As an important result, here using Eqs. (\ref{Q}), (\ref{H,{phi},{quartic},{warm}}), and (\ref{Q,{phi},{quartic},{warm}}), we derive the dissipation parameter in our model as
\begin{equation}
 \label{{Upsilon},{phi},{quartic},{warm}}
 \Upsilon=\frac{\tilde{f}_{0}c_{s}}{\sqrt{f_{0}\left(1-c_{s}^{2}\right)}}\,\phi-\frac{1}{2M_{P}}\sqrt{\frac{3\left(\tilde{f}_{0}c_{s}-4c_{s}+4\right)}{f_{0}c_{s}}}\,\phi^{2}.
\end{equation}
We will discuss about the dissipation parameter (\ref{{Upsilon},{phi},{quartic},{warm}}) in more detail in section \ref{dissipation parameter}.

By solving the differential equation (\ref{{dot}{{phi}},{quartic},{warm}}), we find the evolution of the inflaton field as
\begin{equation}
 \label{{phi},t,{quartic},{warm}}
 \phi=\sqrt{\frac{f_{0}}{1-c_{s}^{2}}}~\frac{1}{t},
\end{equation}
where the integration constant is chosen to be zero without loss of generality. Substituting this into Eqs. (\ref{H,{phi},{quartic},{warm}}), (\ref{Q,{phi},{quartic},{warm}}), and (\ref{T,{phi},{quartic},{warm}}), we can express $H$, $Q$, and $T$, in terms of time, and after inserting these quantities in Eq. (\ref{{mathcal}{P}_s,{warm}}), the scalar power spectrum is obtained as
\begin{equation}
 \label{{mathcal}{P}_s,t,{quartic},{warm}}
 \mathcal{P}_{s}=\frac{f_{0}^{23/8}c_{s}}{192\pi^{3/2}\big(\tilde{f}_{0}c_{s}-4c_{s}+4\big)^{7/8}}\left(\frac{\lambda}{M_{P}t}\right)^{5/2}\left[\frac{2M_{P}\tilde{f}_{0}c_{s}^{3/2}t-\sqrt{3f_{0}\big(\tilde{f}_{0}c_{s}-4c_{s}+4\big)}}{\sqrt{3}\alpha\left(1-c_{s}^{2}\right)^{9}}~\right]^{1/4}.
\end{equation}
Moreover, from Eq. (\ref{{mathcal}{P}_t}) the tensor power spectrum takes the form
\begin{equation}
 \label{{mathcal}{P}_t,t,{quartic},{warm}}
 \mathcal{P}_{t}=\frac{f_{0}\big(\tilde{f}_{0}c_{s}-4c_{s}+4\big)}{6\pi^{2}M_{P}^{4}c_{s}\left(1-c_{s}^{2}\right)^{2}t^{4}}.
\end{equation}
In the following, we want to express these power spectra in terms of the $e$-fold number. We note that if we replace Eq. (\ref{{phi},t,{quartic},{warm}}) into (\ref{H,{phi},{quartic},{warm}}), the Hubble parameter becomes
\begin{equation}
 \label{H,t,{quadratic},{warm}}
 H=\sqrt{\frac{f_{0}\big(\tilde{f}_{0}c_{s}-4c_{s}+4\big)}{3c_{s}}}\frac{1}{2M_{P}\left(1-c_{s}^{2}\right)t^{2}}.
\end{equation}
Using this in Eq. (\ref{{epsilon}}), the first slow-roll parameter is obtained as
\begin{equation}
 \label{{epsilon},t,{quartic},{warm}}
 \epsilon=4M_{P}\big(1-c_{s}^{2}\big)\sqrt{\frac{3c_{s}}{f_{0}\big(\tilde{f}_{0}c_{s}-4c_{s}+4\big)}}~t,
\end{equation}
which apparently increases as time lasts during inflation. Therefore, we can set $\epsilon = 1$ to determine the end time of inflation as
\begin{equation}
 \label{t_e,{quartic},{warm}}
 t_{e}=\frac{1}{4M_{P}\big(1-c_{s}^{2}\big)}\sqrt{\frac{f_{0}\big(\tilde{f}_{0}c_{s}-4c_{s}+4\big)}{3c_{s}}}.
\end{equation}
Here, we are in the position to solve differential equation (\ref{dN}), and using Eqs. (\ref{H,t,{quadratic},{warm}}) and (\ref{t_e,{quartic},{warm}}), we obtain the following relation between the time and the $e$-fold number in our model as
\begin{equation}
 \label{t,N,{quartic},{warm}}
 t=\sqrt{\frac{f_{0}\big(\tilde{f}_{0}c_{s}-4c_{s}+4\big)}{3c_{s}}}\frac{1}{2M_{P}\big(1-c_{s}^{2}\big)\big(N+2\big)}.
\end{equation}
It should be mentioned that to derive the above relation, we have used the initial condition $t\left(N_{e}=0\right)=t_{e}$ in solving the differential equation. Substituting Eq. (\ref{t,N,{quartic},{warm}}) into (\ref{{phi},t,{quartic},{warm}}) gives
\begin{equation}
 \label{{phi},N,{quartic},{warm}}
 \phi=2M_{P}\sqrt{\frac{3c_{s}\left(1-c_{s}^{2}\right)}{c_{s}(\tilde{f}_{0}-4)+4}}\left(N+2\right) \, .
\end{equation}
Replacing this into Eq. (\ref{{Upsilon},{phi},{quartic},{warm}}), one can obtain the dissipation parameter as
\begin{equation}
 \label{{Upsilon},N,{quartic},{warm}}
 \Upsilon=2M_{P}\sqrt{\frac{3c_{s}\lambda}{\tilde{f}_{0}\left[c_{s}(\tilde{f}_{0}-4)+4\right]}}~(N+2)\left[c_{s}\tilde{f}_{0}-3\left(1-c_{s}^{2}\right)(N+2)\right] \, .
\end{equation}
Besides, with the help of Eq. (\ref{t,N,{quartic},{warm}}) in Eq. (\ref{{mathcal}{P}_s,t,{quartic},{warm}}), the scalar power spectrum is given in terms of the number of $e$-foldings as
\begin{equation}
 \label{{mathcal}{P}_s,{quartic},{warm}}
 \mathcal{P}_{s}=\frac{\tilde{f}_{0}^{7/4}\lambda^{3/4}\big(c_{s}(N+2)\big)^{9/4}\left[\tilde{f}_{0}c_{s}-3\big(1-c_{s}^{2}\big)(N+2)\right]^{1/4}}{8\sqrt{2}\pi^{3/2}\alpha^{1/4}\big(\tilde{f}_{0}c_{s}-4c_{s}+4\big)^{2}}.
\end{equation}
Fixing the above expression at the horizon crossing with the $e$-fold number $N_*$, we can readily determine the parameter $\lambda$ as
\begin{equation}
 \label{{lambda},{quartic},{warm}}
 \lambda=\frac{16\times2^{2/3}\pi^{2}\alpha^{1/3}\big(\tilde{f}_{0}c_{s}-4c_{s}+4\big)^{8/3}\mathcal{P}_{s*}^{4/3}}{\tilde{f}_{0}^{7/3}\left(c_{s}\left(N_{*}+2\right)\right)^{3}\left[\tilde{f}_{0}c_{s}-3\left(1-c_{s}^{2}\right)\left(N_{*}+2\right)\right]^{1/3}},
\end{equation}
where the amplitude of the scalar perturbations is constrained as $\ln\left[10^{10}{\cal P}_{s*}\right]=3.094\pm0.034$ according to Planck 2015 TT,TE,EE+lowP data \cite{Planck2015}. According to Eq. (\ref{{lambda},{quartic},{warm}}), it is obvious that for the values  $\tilde{f}_{0}\leq3\left(1-c_{s}^{2}\right)\left(N_{*}+2\right)/c_{s}$ the parameter $\lambda$ becomes imaginary. Therefore, to obtain the real $\lambda$ we will exclude this range of $\tilde{f}_{0}$ in our computations. To calculate the scalar spectral index, we use Eq. (\ref{{mathcal}{P}_s,{quartic},{warm}}) in definition (\ref{n_s}), and also we apply relation (\ref{d{ln}k}). In this way, we reach
\begin{equation}
 \label{n_s,{quartic},{warm}}
 n_{s}=1-\frac{9\tilde{f}_{0}c_{s}-30\left(1-c_{s}^{2}\right)(N+2)}{4(N+2)\left[\tilde{f}_{0}c_{s}-3\left(1-c_{s}^{2}\right)(N+2)\right]}.
\end{equation}
If we apply relation (\ref{d{ln}k}) again in the above equation, the running of the scalar spectral index will be resulted in as
\begin{equation}
 \label{dn_s/d{ln}k,{quartic},{warm}}
 \frac{dn_{s}}{d\ln k}=-\frac{9\left[10\big(1+c_{s}^{4}\big)(N+2)^{2}-6\tilde{f}_{0}c_{s}\big(1-c_{s}^{2}\big)(N+2)-c_{s}^{2}\big(20N^{2}+80N-\tilde{f}_{0}^{2}+80\big)\right]}{4(N+2)^{2}\left[\tilde{f}_{0}c_{s}-3\left(1-c_{s}^{2}\right)(N+2)\right]^{2}}.
\end{equation}

In order to find the tensor power spectrum of our model in terms of the $e$-fold number, we substitute Eq. (\ref{t,N,{quartic},{warm}}) into  (\ref{{mathcal}{P}_t,t,{quartic},{warm}}), and obtain
\begin{equation}
 \label{{mathcal}{P}_t,{quartic},{warm}}
 \mathcal{P}_{t}=\frac{24\lambda c_{s}\left(1-c_{s}^{2}\right)^{2}(N+2)^{4}}{\pi^{2}\tilde{f}_{0}\big(\tilde{f}_{0}c_{s}-4c_{s}+4\big)}.
\end{equation}
Therefore, by using this together with Eq. (\ref{d{ln}k}), it is simple to show that definition (\ref{n_t}) leads to
\begin{equation}
 \label{n_t,{quartic},{warm}}
 n_{t}=-\frac{4}{N+2}.
\end{equation}
Furthermore, by replacing Eqs. (\ref{{mathcal}{P}_s,{quartic},{warm}}) and (\ref{{mathcal}{P}_t,{quartic},{warm}}) into definition (\ref{r}), and after applying Eq. (\ref{{lambda},{quartic},{warm}}), it is straightforward to derive the tensor-to-scalar ratio as
\begin{equation}
 \label{r,{quartic},{warm}}
 r=\frac{384\left(1-c_{s}^{2}\right)^{2}(N+2)}{c_{s}^{2}}\left(\frac{4\alpha\big(\tilde{f}_{0}c_{s}-4c_{s}+4\big)^{5}\mathcal{P}_{s*}}{\tilde{f}_{0}^{10}\left[\tilde{f}_{0}c_{s}-3\big(1-c_{s}^{2}\big)(N+2)\right]}\right)^{1/3}.
\end{equation}
Now, we can plot the $r-n_s$ diagram for our model by the use of Eqs. (\ref{n_s,{quartic},{warm}}) and (\ref{r,{quartic},{warm}}). This plot is illustrated in Fig. \ref{figure:r,n_s} for some typical values of the sound speed $c_s$. In the figure, the dashed and solid lines correspond to $N_*=50$ and $N_*=60$, respectively. Figure \ref{figure:r,n_s} shows that, the $r-n_s$ prediction of the quartic potential (\ref{V,{quartic}}) can enter even the 68\% CL region favored by Planck 2015 TT,TE,EE+lowP data \cite{Planck2015} in the warm DBI inflationary setting, whereas its result in the cold DBI setting, is completely ruled out by the observational data. Here, we take $\tilde{f}_{0}$ as the varying parameter in the range of $\tilde{f}_{0} > 3\left(1-c_{s}^{2}\right)\left(N_{*}+2\right)/c_{s}$ to plot the $r-n_s$ diagram of this model, such that it increases from top to down of each plot. Note that we exclude the values $\tilde{f}_{0} \leq 3\left(1-c_{s}^{2}\right)\left(N_{*}+2\right)/c_{s}$ from our computations, because as we already mentioned they give rise to imaginary values for $\lambda$. In Table \ref{table:warm_DBI}, we have presented the ranges of $\tilde{f}_{0}$ for which our model with several values of $c_s$ and with $N_*=50,\,60$, verifies the 68\% CL constraints of Planck 2015 TT,TE,EE+lowP data \cite{Planck2015} in the $r-n_s$ plane.

\begin{figure}[t]
\begin{center}
\scalebox{1}[1]{\includegraphics{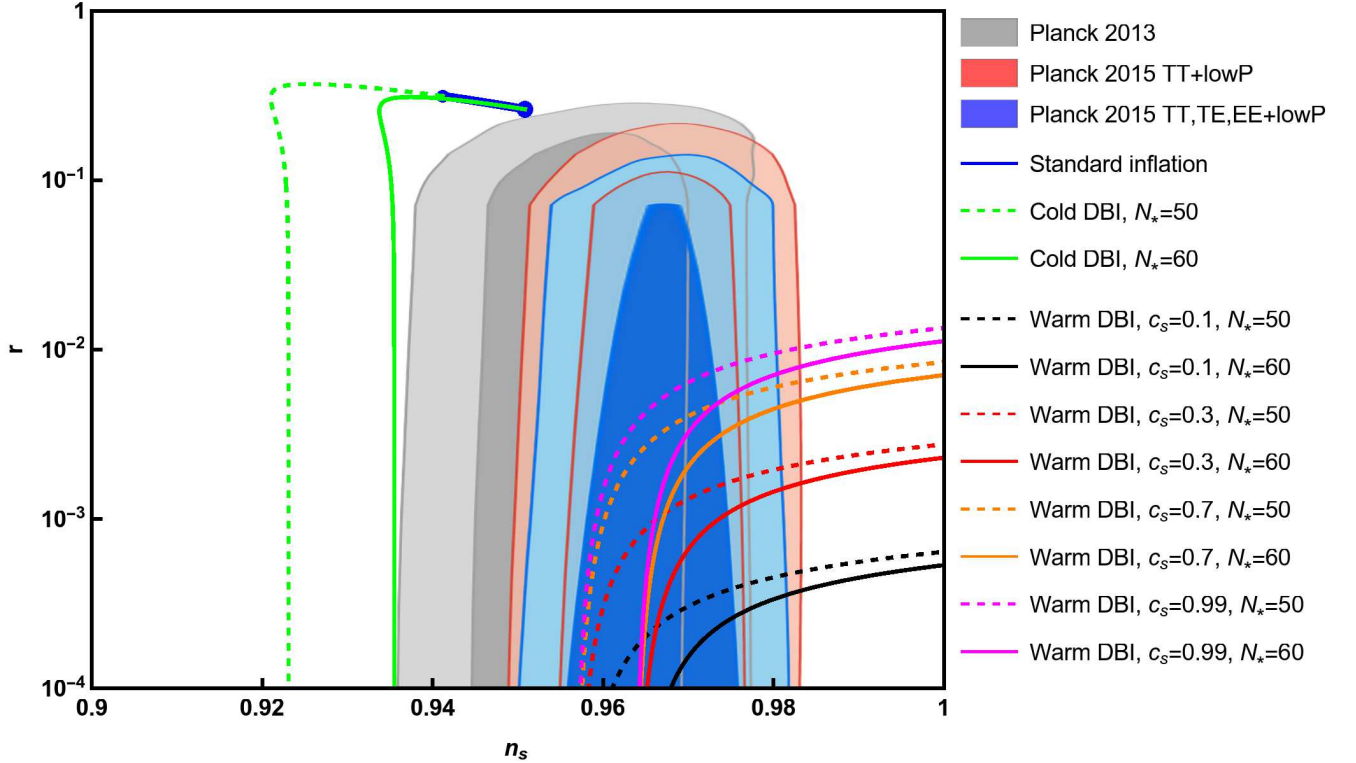}}
\caption{The $r-n_s$ diagram of the quartic potential (\ref{V,{quartic}}) in the cold and warm DBI inflationary settings. The dashed and solid lines are related to $N_*=50$ and $N_*=60$, respectively. The marginalized joint 68\% and 95\% CL regions of Planck 2013, Planck 2015 TT+lowP, and Planck 2015 TT,TE,EE+lowP data \cite{Planck2015} are specified by gray, red, and blue, respectively.}
\label{figure:r,n_s}
\end{center}
\end{figure}

\begin{table}
 \centering
 \caption{The ranges of parameter $\tilde{f}_{0}$ for which our warm DBI inflation model with several values of the constant sound speed $c_s$ and with $N_*=50,\,60$, verifies the 68\% CL constraints of Planck 2015 TT,TE,EE+lowP data \citep{Planck2015} in the $r-n_s$ plane.}
 \scalebox{1}{
 \begin{tabular}{|c|c|c|}
  \hline
  $\quad$ $c_{s}$ $\quad$ & $\quad$ $N_{*}$ $\quad$ & $\qquad\qquad$ $\tilde{f}_{0}$ $\qquad\qquad$ \tabularnewline
  \hline
  \hline
  \multirow{2}{*}{0.1} & 50 & $\tilde{f}_{0}\gtrsim1.95\times10^{3}$\tabularnewline
  \cline{2-3}
  & 60 & $\tilde{f}_{0}\gtrsim2.48\times10^{3}$\tabularnewline
  \hline
  \multirow{2}{*}{0.3} & 50 & $\tilde{f}_{0}\gtrsim6.03\times10^{3}$\tabularnewline
  \cline{2-3}
  & 60 & $\tilde{f}_{0}\gtrsim7.75\times10^{3}$\tabularnewline
  \hline
  \multirow{2}{*}{0.7} & 50 & $\tilde{f}_{0}\gtrsim1.46\times10^{2}$\tabularnewline
  \cline{2-3}
  & 60 & $\tilde{f}_{0}\gtrsim1.90\times10^{2}$\tabularnewline
  \hline
  \multirow{2}{*}{0.99} & 50 & $\tilde{f}_{0}\gtrsim4.07$\tabularnewline
  \cline{2-3}
  & 60 & $\tilde{f}_{0}\gtrsim5.30$\tabularnewline
  \hline
 \end{tabular}
 }
\label{table:warm_DBI}
\end{table}

In Fig. \ref{figure:dn_s/d{ln}k,n_s,{quartic},{warm}}, we have used Eqs. (\ref{n_s,{quartic},{warm}}) and (\ref{dn_s/d{ln}k,{quartic},{warm}}) to draw the prediction of our warm DBI inflation model with $c_s=0.1$ in the $dn_s/d\ln k-n_s$ plane. The dashed and solid black lines show the results of our model with $N_*=50$ and $N_*=60$, respectively, and for the both cases, our model is compatible with the 68\% CL constraints of Planck 2015 TT,TE,EE+lowP data \cite{Planck2015}. Notice that to draw the plot, we take again $\tilde{f}_{0} > 3\left(1-c_{s}^{2}\right)\left(N_{*}+2\right)/c_{s}$ which provides the real values for $\lambda$ in Eq. (\ref{{lambda},{quartic},{warm}}).

\begin{figure}[t]
\begin{center}
\scalebox{1}[1]{\includegraphics{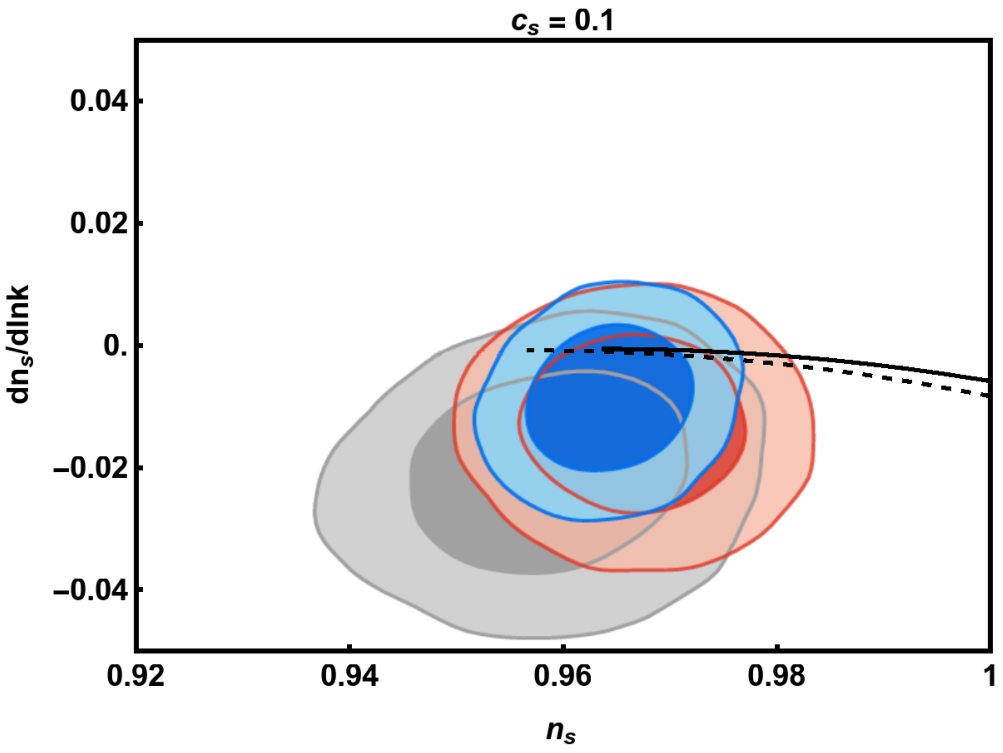}}
\caption{The $dn_s/d\ln k-n_s$ diagram of the quartic potential (\ref{V,{quartic}}) in the warm DBI inflationary setting with $c_s=0.1$. The dashed and solid lines correspond to $N_*=50$ and $N_*=60$, respectively. The marginalized joint 68\% and 95\% CL regions of Planck 2013, Planck 2015 TT+lowP, and Planck 2015 TT,TE,EE+lowP data \cite{Planck2015} are specified by gray, red, and blue, respectively.}
\label{figure:dn_s/d{ln}k,n_s,{quartic},{warm}}
\end{center}
\end{figure}

It is also useful to examine the behavior of the inflationary observables in the limit $\tilde{f}_{0}\gg1$. In this limit, Eq. (\ref{{mathcal}{P}_s,{quartic},{warm}}) turns into
\begin{equation}
 \label{{mathcal}{P}_s,{quuartic},{warm},{tilde}{f}_0{gg}1}
 \mathcal{P}_{s}=\frac{1}{8}\sqrt{\frac{c_{s}}{2\pi^{3}}}
 \left(\frac{\lambda^{3}(N+2)^{9}}{\alpha}\right)^{1/4},
\end{equation}
where at the horizon exit $e$-fold number $N_*$, the solution of Eq. (\ref{{mathcal}{P}_s,{quuartic},{warm},{tilde}{f}_0{gg}1}) for $\lambda$ is
\begin{equation}
 \label{{lambda},{warm},{quartic},{tilde}{f}_0{gg}1}
 \lambda=\frac{16\pi^{2}}{\left(N_{*}+2\right)^{3}}\left(\frac{4\alpha\mathcal{P}_{s*}^{4}}{c_{s}^{2}}\right)^{1/3}.
\end{equation}
Additionally, Eqs. (\ref{n_s,{quartic},{warm}}), (\ref{dn_s/d{ln}k,{quartic},{warm}}), and (\ref{r,{quartic},{warm}}) are simplified in the limit $\tilde{f}_{0}\gg1$ to
\begin{align}
 \label{n_s,{quartic},{warm},{tilde}{f}_0{gg}1}
 n_{s} &=\frac{4N-1}{4(N+2)},
 \\
 \label{dn_s/d{ln}k,{quartic},{warm},{tilde}{f}_0{gg}1}
 \frac{dn_{s}}{d\ln k} &=-\frac{9}{4(N+2)^{2}},
 \\
 \label{r,{quartic},{warm},{tilde}{f}_0{gg}1}
 r &=0.
\end{align}
It is important to note that the above equations are independent of the parameter $c_s$. Duo to this fact, the $r-n_s$ plots in Fig. \ref{figure:r,n_s} tend to the same results for a given $N_{*}$ when the parameter $\tilde{f}_{0}$ goes infinity. In this limit from Eq. (\ref{n_s,{quartic},{warm},{tilde}{f}_0{gg}1}) for $N_*=50$ and $60$, respectively, we obtain $n_s=0.95673$ and $0.9637$, while from Eq. (\ref{r,{quartic},{warm},{tilde}{f}_0{gg}1}) we have $r=0$ for the both cases. These results are in excellent agreement with Planck 2015 TT,TE,EE+lowP data \cite{Planck2015}, since they lie perfectly inside the 68\% CL constraints of these data set in $r-n_s$ plane. Also, for $N_*=50$ and $60$, Eq. (\ref{dn_s/d{ln}k,{quartic},{warm},{tilde}{f}_0{gg}1}) results in $dn_s/d\ln k=-0.0008$ and $-0.0006$, respectively, which are compatible with the 95\% CL constraint provided by Planck 2015 TT,TE,EE+lowP data combination \cite{Planck2015}.

In the next step, by the use of Eqs. (\ref{n_s,{quartic},{warm}}) and (\ref{r,{quartic},{warm}}) again, we have specified the parameter space of $\tilde{f}_{0}-c_{s}$ in Figs. \ref{figure:{tilde}{f}_0,c_s,N_*=50} and \ref{figure:{tilde}{f}_0,c_s,N_*=60} for $N_*=50$ and $60$, respectively. In these figures, the blue region points out the parameter space for which the $r-n_s$ prediction of our model is compatible with the 68\% CL constraints of Planck 2015 TT,TE,EE+lowP data \cite{Planck2015}. We deduce from these figures that our model is compatible with the observations provided that $\tilde{f}_0\gtrsim 10^2$. In these figures, we have also have separated the region $T>H$ from the region $T<H$ by a black curve. In fact, we have drawn the black curves in the figures to check the basic condition $T>H$ in our warm DBI inflationary model. To draw these curves, we have used the following equation
\begin{equation}
 \label{T/H,{quadratic},{warm}}
 \frac{T}{H}=\frac{1}{\sqrt{\pi c_{s}\left(1-c_{s}^{2}\right)}}\left[\frac{6\tilde{f}_{0}c_{s}\left[\tilde{f}_{0}c_{s}-3\left(1-c_{s}^{2}\right)(N+2)\right]}{223\lambda(N+2)^{5}}\right]^{1/4},
\end{equation}
which follows from applying Eqs. (\ref{{phi},t,{quartic},{warm}}) and (\ref{t,N,{quartic},{warm}}) in Eqs. (\ref{H,{phi},{quartic},{warm}}) and (\ref{T,{phi},{quartic},{warm}}). Solving $T/H=1$ in Eq. (\ref{T/H,{quadratic},{warm}}) leads to the black curves in Figs. \ref{figure:{tilde}{f}_0,c_s,N_*=50} and \ref{figure:{tilde}{f}_0,c_s,N_*=60}. As we see in the these figures, the compatibility region of our model lies on the top of the black curve which is related to $T>H$. Therefore, our model satisfies the constraints of the Planck 2015 results without disturbing the essential condition $T>H$ of warm inflation. In order to specify the different regimes of warm inflation, we have drawn the dashed green and orange plots which are associated to $Q_*=0.1$ and $Q_*=10$, respectively. To draw these curves, we have used
\begin{equation}
 \label{Q,N,{quartic},{warm}}
 Q=\frac{\tilde{f}_{0}c_{s}}{3\left(1-c_{s}^{2}\right)(N+2)}-1,
\end{equation}
which has been obtained from substituting Eqs. (\ref{{phi},t,{quartic},{warm}}) and (\ref{t,N,{quartic},{warm}}) into (\ref{Q,{phi},{quartic},{warm}}). From the dashed green and orange curves, we conclude that our warm DBI inflation model is consistent with the Planck 2015 results in the intermediate ($0.1\lesssim Q_*\lesssim10$) and strong ($Q_*\gtrsim10$) dissipation regimes, and it cannot compatible with the cosmological data in the weak ($Q_*\lesssim0.1$) dissipation regime. However, consideration of the modifications duo to the high temperatures regime which change the distribution of the inflaton particles from the vacuum phase space state to the excited Bose-Einstein state \cite{Bartrum2014, Bastero-Gil2016}, may affect the results of the model in hand, so that it can be compatible with the observational data even in the weak dissipation regime, and we leave this idea to the future investigations.

\begin{figure}[t]
\begin{center}
\scalebox{1}[1]{\includegraphics{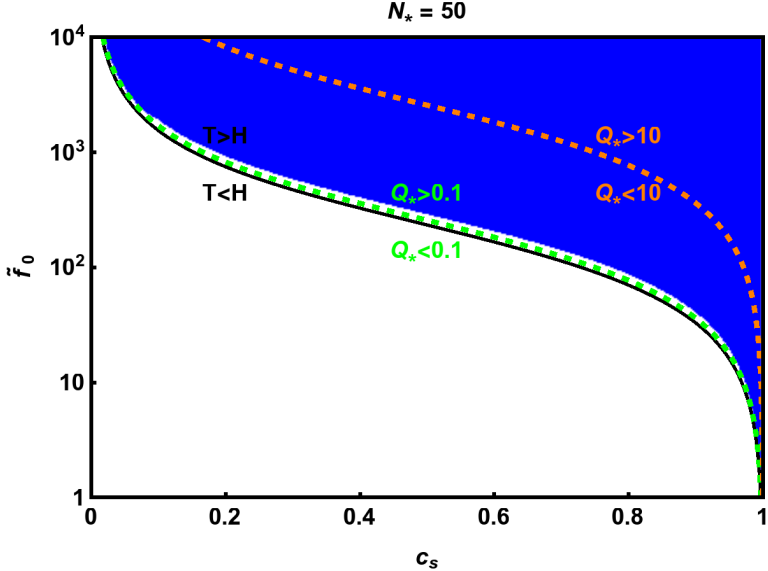}}
\caption{The parameter space of $\tilde{f}_{0}-c_{s}$ for which our warm DBI inflationary model with $N_*=50$ is compatible with the 68\% CL joint region of Planck 2015 TT,TE,EE+lowP data \cite{Planck2015} in $r-n_s$ plane. The dashed green and orange curves separate the weak ($Q_*=0.1$) and strong ($Q_*=10$) dissipation regimes.}
\label{figure:{tilde}{f}_0,c_s,N_*=50}
\end{center}
\end{figure}

\begin{figure}
\begin{center}
\scalebox{1}[1]{\includegraphics{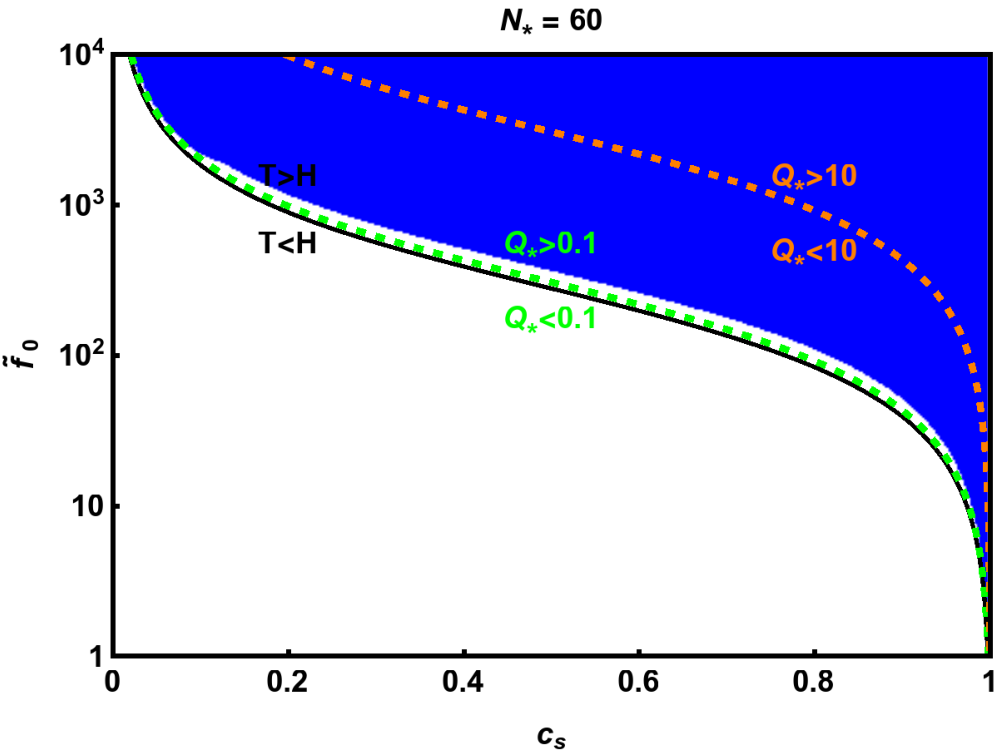}}
\caption{Same as Fig. \ref{figure:{tilde}{f}_0,c_s,N_*=50}, but for $N_*=60$.}
\label{figure:{tilde}{f}_0,c_s,N_*=60}
\end{center}
\end{figure}

We further can characterize the domain of the parameter $\lambda$ of the quartic potential (\ref{V,{quartic}}) for which our inflation model fulfills the Planck 2015 observational data constraints. In Eq. (\ref{{lambda},{quartic},{warm}}) we obtained a relation for $\lambda$ which follows from the fixing of the scalar power spectrum at the time of horizon crossing. Here, we use Eq. (\ref{{lambda},{quartic},{warm}}) together with Eqs. (\ref{n_s,{quartic},{warm}}) and (\ref{r,{quartic},{warm}}) to plot the parameter space of $\lambda-c_s$ for which our warm DBI inflation is in agreement with the 68\% CL region of Planck 2015 TT,TE,EE+lowP data \cite{Planck2015} in $r-n_s$ plane. The $\lambda-c_s$ diagrams of model with $N_*=50$ and $N_*=60$ are displayed in Figs. \ref{figure:{lambda},c_s,{quartic},{warm},N_*=50} and \ref{figure:{lambda},c_s,{quartic},{warm},N_*=60}, respectively. These figures reflects the point that our model is compatible with the current data provided $\lambda\lesssim10^{-13}$, which is in explicit contrast with the bound $\lambda\simeq0.13$ for the Higgs coupling implied by the LHC measurements \cite{ATLAS2012, CMS2012}. However, it should be reminded that it is not necessary that this experimental bound is preserved in our model, because here the role of inflaton is accomplished by the radial coordinate of a D3-brane \cite{Silverstein2004, Alishahiha2004}, not by the Higgs boson of the standard particle physics. Combining the constraint $\lambda\lesssim10^{-13}$ with the bound $\tilde{f}_0\gtrsim 10^2$ deduced from the $\tilde{f}_0-c_s$ parameter space, we find the lower bound $f_0\gtrsim 10^{15}$ in our warm DBI inflation model.

\begin{figure}[t]
\begin{center}
\scalebox{1}[1]{\includegraphics{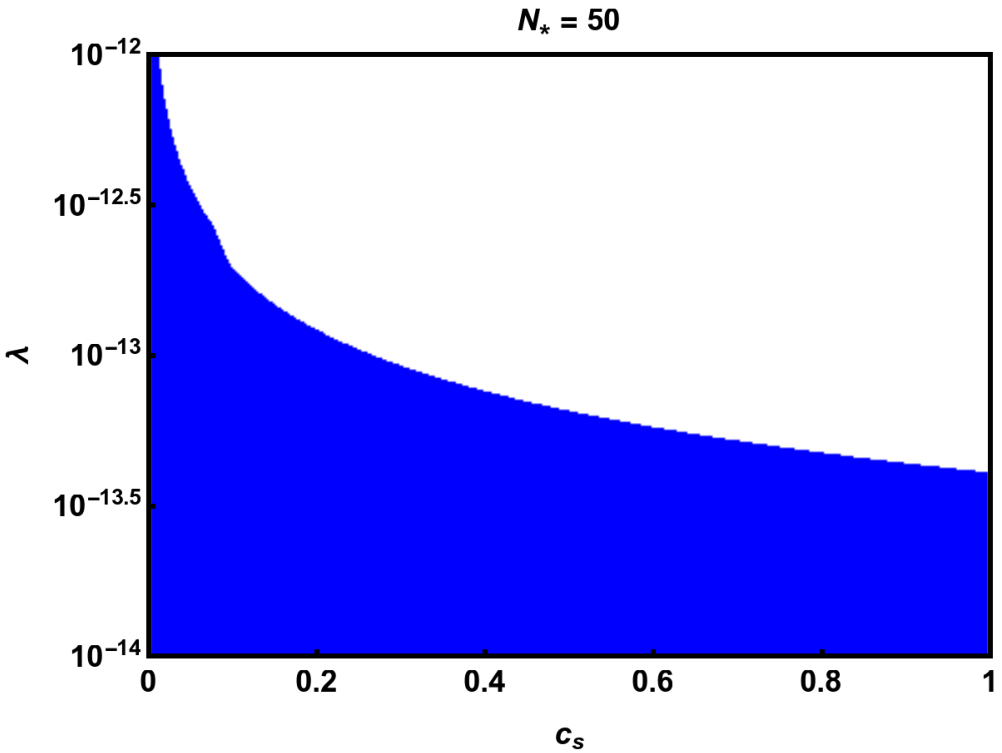}}
\caption{The parameter space of $\lambda-c_{s}$ for which our warm DBI inflationary model with $N_*=50$ is compatible with the 68\% CL joint region of Planck 2015 TT,TE,EE+lowP data \cite{Planck2015} in $r-n_s$ plane.}
\label{figure:{lambda},c_s,{quartic},{warm},N_*=50}
\end{center}
\end{figure}

\begin{figure}
\begin{center}
\scalebox{1}[1]{\includegraphics{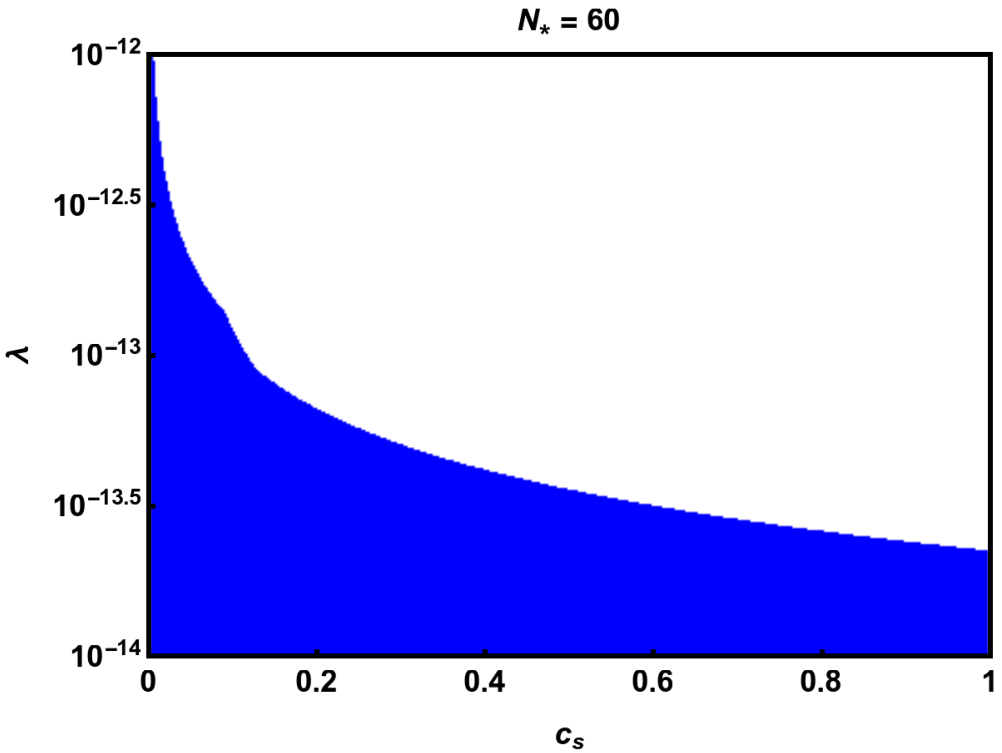}}
\caption{Same as Fig. \ref{figure:{lambda},c_s,{quartic},{warm},N_*=50}, but for $N_*=60$.}
\label{figure:{lambda},c_s,{quartic},{warm},N_*=60}
\end{center}
\end{figure}


\section{Some regards on the dissipation parameter}
\label{dissipation parameter}

Here, we discuss in more detail about the dissipation parameter (\ref{{Upsilon},{phi},{quartic},{warm}}) in our warm inflationary model. For this purpose, we rewrite Eq. (\ref{{Upsilon},{phi},{quartic},{warm}}) as
\begin{equation}
 \label{{Upsilon},{Upsilon}_1,{Upsilon}_2}
 \Upsilon=\Upsilon_{1}+\Upsilon_{2} \, ,
\end{equation}
where we have defined
\begin{align}
 \label{{Upsilon}_1}
 \Upsilon_{1} & \equiv\frac{c_{s}\tilde{f}_{0}}{\sqrt{f_{0}\left(1-c_{s}^{2}\right)}}\,\phi \, ,
 \\
 \label{{Upsilon}_2}
 \Upsilon_{2} & \equiv-\frac{1}{2M_{P}}\sqrt{\frac{3\left(c_{s}\tilde{f}_{0}-4c_{s}+4\right)}{f_{0}c_{s}}}\,\phi^{2} \, .
\end{align}
If we calculate the absolute value of the ratio of $\Upsilon_{2}$ to $\Upsilon_{1}$, and also use Eq. (\ref{{phi},N,{quartic},{warm}}), we will have
\begin{equation}
 \label{{Upsilon}_2/{Upsilon}_1}
 \left|\frac{\Upsilon_{2}}{\Upsilon_{1}}\right|=\frac{3\left(1-c_{s}^{2}\right)(N+2)}{c_{s}\tilde{f}_{0}} \, .
\end{equation}
We evaluate this quantity at the horizon crossing with $N_*=(50, \, 60)$ for $c_s=(0.1, \, 0.3, \, 0.7, \, 0.99)$ while the parameter $\tilde{f}_0$ is chosen in the ranges presented in Table \ref{table:warm_DBI} for which our model is compatible with the Planck 2015 data. The plots are presented in Fig. \ref{figure:{Upsilon}_2/{Upsilon}_1}, and we see that $\left|\frac{\Upsilon_{2}}{\Upsilon_{1}}\right|$ is much smaller than unity. Therefore, the term $\Upsilon_{2}$ in the dissipation parameter can be neglected in comparison with $\Upsilon_{1}$, and consequently the dissipation parameter behaves approximately as $\Upsilon\propto\phi$. Besides, we can also prove this fact analytically in the limit $\tilde{f}_0 \gg 1$. In this limit, Eq. (\ref{{Upsilon},N,{quartic},{warm}}) is simplified to
\begin{equation}
 \label{{Upsilon},N,{quartic},{warm},{tilde}{f}_0{gg}1}
 \Upsilon=2M_{P}c_{s}\sqrt{3\lambda}(N+2) \, .
\end{equation}
If we compare this equation with Eq. (\ref{{phi},N,{quartic},{warm}}), we see that
\begin{equation}
 \label{{Upsilon},{phi},{tilde}{f}_0{gg}1}
 \Upsilon=C_{\phi}\,\phi \, ,
\end{equation}
where the dissipation constant $C_{\phi}$ is given by
\begin{equation}
 \label{C_{phi}}
 C_{\phi}=\sqrt{\frac{c_{s}\lambda\left(c_{s}\tilde{f}_{0}-4c_{s}+4\right)}{1-c_{s}^{2}}} \, .
\end{equation}
Therefore, in our DBI warm inflationary model and in the limit $\tilde{f}_0 \gg 1$, the dissipation parameter behaves as $\Upsilon\propto\phi$ which is in agreement with the numerical results presented in Fig. \ref{figure:{Upsilon}_2/{Upsilon}_1}.

Note that the dissipation parameter $\Upsilon\propto\phi$ is also supported by QFT as well as phenomenological implications. For instance, the authors of \cite{Moss2006} by applying the superpotential $W=\frac{g}{\sqrt{2}}\Phi Y^{2}-\frac{h}{\sqrt{2}}YZ^{2}$ in a supersymmetric theory, and by considering an exponential decay approximation for the propagator in the quantum field theory (QFT) interactions, found \cite{Moss2006}
\begin{equation}
 \label{Upsilon_{MS}}
 \Upsilon_{\rm MS}=\frac{g^{3}h^{2}}{16\pi^{2}}\phi\left(2+\frac{g\phi}{m_{y}}+\frac{g^{3}\phi^{3}}{m_{y}^{3}}\right) \, .
\end{equation}
The subscript $MS$ refers to Morikawa and Sasaki \cite{Morikawa1984} who found for the first time a similar result for non-supersymmetric models, and their result was verified later by Berera and Ramos \cite{Berera2001, Berera2003}. For a special case, if we have $m_{y}\gg\left|g\phi\right|$, then the second and third terms in the bracket in Eq. (\ref{Upsilon_{MS}}) can be ignored against the first term, and the dissipation parameter turns into $\Upsilon\propto\phi$. This form for the dissipation parameter is also interesting at the phenomenological point of view. In \cite{Zhang2009}, a general form $\Upsilon=C_{\phi}\frac{T^{m}}{\phi^{m-1}}$ was proposed phenomenologically for the dissipation parameter, where the constant $C_\phi$ depends on the microphysics of the dissipation process, and the exponent $m$ is an integer. The special cases of this dissipation parameter  can also be extracted from the interactions given by the superpotential $W$. One special case is $m=1$ which gives $\Upsilon\propto T$ that corresponds to the high temperature supersymmetry model \cite{Moss2006}. Also, other special cases are $m=0$ ($\Upsilon\propto\phi$) and $m=-1$ ($\Upsilon\propto \phi^2/T$) which are related to an exponentially decaying propagator in the high temperature supersymmetric and non-supersymmetric models, respectively \cite{Berera2001, Yokoyama1999}. Some further investigations on the dissipation parameter $\Upsilon\propto\phi$ in the framework of warm inflation can be found in \cite{Berera2005, Hall2004, Hall2005, Hall2008, Zhang2009}.

\begin{figure}[t]
\begin{center}
\scalebox{0.45}[0.45]{\includegraphics{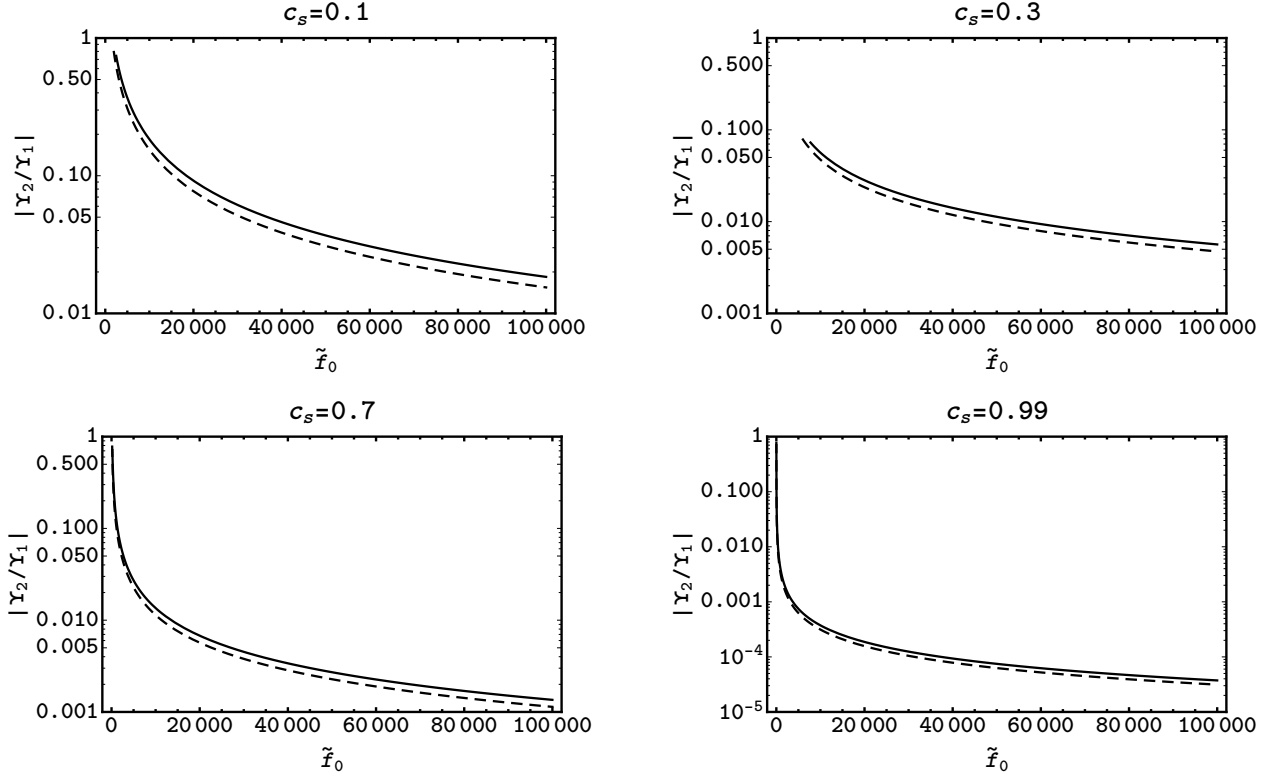}}
\caption{The variations of $\left|\Upsilon_{2}/\Upsilon_{1}\right|$ versus $\tilde{f}_0$ for different sound speeds $c_s=(0.1, \, 0.3, \, 0.7, \, 0.99)$. The dashed and solid lines correspond to $N_*=$ 50 and 60, respectively.}
\label{figure:{Upsilon}_2/{Upsilon}_1}
\end{center}
\end{figure}


\section{Examination of the swampland conjecture}
\label{section:swampland}

\begin{figure}[t]
\begin{center}
\scalebox{0.45}[0.45]{\includegraphics{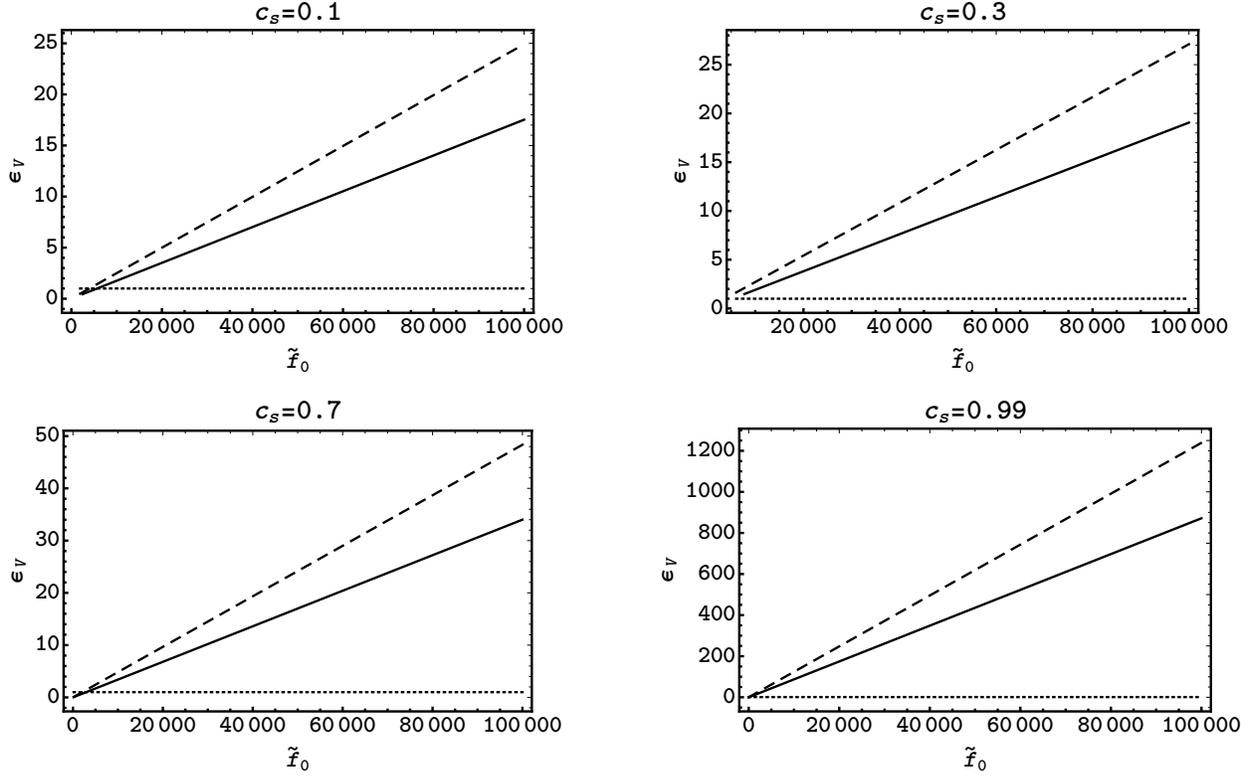}}
\caption{The potential slow-roll parameter $\epsilon_V$ versus $\tilde{f}_0$ for different sound speeds $c_s=(0.1, \, 0.3, \, 0.7, \, 0.99)$. The dashed and solid lines are related to $N_*=$ 50 and 60, respectively.}
\label{swampland}
\end{center}
\end{figure}

In this section, we proceed to examine the swampland criteria \cite{Agrawal2018, Ooguri2018} in our warm DBI inflationary model. This problem has recently been proposed in the string theory and it challenges the standard single field inflation. The refined version of the swampland criterion implies that to satisfy the dS limit in the standard inflation, it is necessary that \cite{Ooguri2018}
\begin{equation}
 \label{{nabla}V}
 \left|\nabla V\right|\geq\frac{c}{M_{P}}\,V,
\end{equation}
or
\begin{equation}
 \label{{nabla}_i{nabla}_jV}
 \min\left(\nabla_{i}\nabla_{j}V\right)\leq-\frac{c'}{M_{P}^{2}}\,V,
\end{equation}
where $V$ is the scalar field potential and $c, \, c'>0$ are some universal constants of order 1. The left hand side of Eq. (\ref{{nabla}_i{nabla}_jV}) is the minimum of the Hessian $\nabla_{i}\nabla_{j}V$ in an orthonormal frame. In the standard supercold inflation, validity of the criteria (\ref{{nabla}V}) and (\ref{{nabla}_i{nabla}_jV}) is in contrast with the slow-roll conditions, and therefore we encounter the so-called swampland problem. But, the authors of \cite{Motaharfar2018, Das2018} have shown that the validity of the swampland criteria does not lead to any contradiction in the setup of warm inflation. Here, we are also interested in examining the validity of the swampland criteria in our warm DBI inflation model. To this aim, we first introduce the potential slow-roll parameter that is conventionally defined as
\begin{equation}
 \label{{epsilon}_V}
 \epsilon_{V}\equiv\frac{M_{P}^{2}}{2}\left(\frac{V'(\phi)}{V(\phi)}\right)^{2} ~ .
\end{equation}
This parameter measures the slope of inflationary potential and in order to satisfy the swampland creation (\ref{{nabla}V}), it must be of order of unity or larger (i.e. $\epsilon_V \gtrsim 1$). In the standard cold inflation, this parameter has relation with the Hubble slow-roll parameter as $\epsilon_V \approx \epsilon\equiv-\dot{H}/H^2$, and therefore the slow-roll condition $\epsilon \ll 1$ is inconsistent with the dS requirement $\epsilon_V \gtrsim 1$. But in the framework of warm DBI inflation, if we use the slow-roll equations (\ref{H,{rho}_{phi},{warm}}) and (\ref{{dot}{{phi}},{warm}}), we can simply show that the potential slow-roll parameter (\ref{{epsilon}_V}) is related to the Hubble slow-roll parameter (\ref{{epsilon}}) as
\begin{equation}
 \label{{epsilon}_V,{epsilon}}
 \epsilon_{V}\approx\frac{(1+Q)}{c_{s}}~\epsilon.
\end{equation}
This relation implies that in the warm DBI inflation the dS requirement $\epsilon_V \gtrsim 1$ can be preserved even if $\epsilon \ll 1$ provided that the dissipation ratio $Q$ is chosen large enough. In order to show this fact in our investigation more explicitly, we calculate the potential slow-roll parameter (\ref{{epsilon}_V}) by the use of Eqs. (\ref{V,{quartic}}) and (\ref{{phi},N,{quartic},{warm}}), and reach
\begin{equation}
 \label{{epsilon}_V,f0tilde}
 \epsilon_{V}=\frac{2\left[c_{s}\left(\tilde{f}_{0}-4\right)+4\right]}{3c_{s}\left(1-c_{s}^{2}\right)\left(N+2\right)^{2}} ~ .
\end{equation}
In Fig. \ref{swampland}, we plot $\epsilon_V$, Eq. (\ref{{epsilon}_V,f0tilde}), versus $\tilde{f}_0$ for $c_s=(0.1, \, 0.3, \, 0.7, \, 0.99)$ and $N_*=(50, \, 60)$. To draw these plots, the range of the parameter $\tilde{f}_0$ is varying according to Table \ref{table:warm_DBI} for which our model is compatible with the Planck 2015 data. As we see in the figure, the dS requirement $\epsilon_V \gtrsim 1$ is satisfied. This confirms that the swampland problem can be resolved in our warm DBI inflation model. Here, one should note that to overcome the swampland problem, one of the criteria (\ref{{nabla}V}) or (\ref{{nabla}_i{nabla}_jV}) is enough to be respected. However, we checked numerically that the second swampland creation (\ref{{nabla}_i{nabla}_jV}) in contrast to the first one (\ref{{nabla}V}) is not preserved in our model.

\section{Conclusions}
\label{section:conclusions}

The DBI inflation has well-based motivations from the sting theory, and in this model the role of inflaton is played by a radial coordinate of a D3-brane moving in a throat of a compactification space. Therefore, this scenario proposes a convincing candidate for the inflaton field. In addition, thus far some suggestions have been offered for the inflationary potential in the DBI inflation which can alleviate the eta problem theoretically in this setting relative to the conventional inflationary frameworks. In this paper, we study the DBI inflation in both the cold and warm scenarios. We first focused on the cold DBI inflation and reviewed shortly the basic equations of the background dynamics and the scalar and tensor primordial perturbations in this setting. At the level of nonlinear perturbations, we showed that the 95\% CL observational constraint of Planck 2015 T+E data \cite{Planck2015non-Gaussianity} on the primordial non-Gaussianity leads to the bound $c_{s}\geq0.087$ on the sound speed of the model. Then, we investigated the quartic potential $V(\phi)=\lambda\phi^{4}/4$ in the context of cold DBI inflation with the AdS warp factor $f(\phi)=f_{0}/\phi^{4}$, and checked its viability in light of the Planck 2015 results. Our examination implied that the result of the quartic potential in the cold DBI setting like the standard scenario is completely outside the allowed region of  Planck 2015 TT,TE,EE+lowP data \cite{Planck2015} in $r-n_s$ plane.

Subsequently, we turned to study the warm DBI inflation, and presented the basic equations of this model. In this framework, the nonlinear perturbations of the inflaton dominates over the thermal contribution, provided that the sound speed be small enough. Consequently, in this scenario we recover the same lower bound found in the cold DBI inflation for the sound speed from the 95\% CL constraint of Planck 2015 T+E data on the primordial non-Gaussianity parameter.

In the warm DBI inflation, we examined the same quartic potential $V(\phi)=\lambda\phi^{4}/4$ with the AdS warp factor $f(\phi)=f_{0}/\phi^{4}$ and assumed the sound speed to be a constant quantity during inflation. In the next step, we displayed the result of our model in $r-n_s$ plane in comparison with the observational results. We saw that our model with different values of $c_s$ is consistent with the CMB data, and its prediction can enter within the 68\% CL region of Planck 2015 TT,TE,EE+lowP data. We also demonstrated that the $dn_s/d\ln k-n_s$ diagram of our warm DBI inflationary model with $c_s = 0.1$ can be located inside the 68\% CL region favored by Planck 2015 TT,TE,EE+lowP data.

 We further specified the parameter space of $\tilde{f}_{0}-c_{s}$ ($\tilde{f}_{0}\equiv f_{0}\lambda$) for which the $r-n_s$ prediction of our model is consistent with the 68\% CL constraints of Planck 2015 TT,TE,EE+lowP data. This parameter space implies that our model is compatible with the observations provided that $\tilde{f}_0\gtrsim 10^2$. Using the parameter space of $\tilde{f}_{0}-c_{s}$, we also showed that the essential requirement of warm inflation, i.e. $T>H$, is perfectly preserved in our model. From this parameter space, we also concluded that our model can be compatible with observation in the intermediate ($0.1\lesssim Q_*\lesssim10$) and high ($Q_*\gtrsim10$) dissipation regimes of warm inflation, and it fails to be consistent in the weak ($Q_*\lesssim0.1$) dissipation regime. Nevertheless, if one consider the modifications duo to the high temperatures regime which alter the distribution of the inflaton particles from the vacuum phase space state to the excited Bose-Einstein state \cite{Bartrum2014, Bastero-Gil2016}, the results of this model may be compatible with the observational data even in the weak dissipation regime, and we leave this suggestion to the future investigations.

Furthermore, in order to estimate the value of $\lambda$ following from fixing the scalar power spectrum at the horizon exit in our model, we presented the parameter space of $\lambda-c_s$, and from it we concluded that our model is in consistency with the Planck 2015 results provided $\lambda\lesssim10^{-13}$. This is in direct contrast with the value $\lambda\simeq0.13$ measured by the LHC at CERN for the coupling constant of the Higgs boson of the standard model of particle physics. Nonetheless, it is not required that this bound is kept in our model, because here the role of inflaton is performed by the radial coordinate of a D3-brane \cite{Silverstein2004, Alishahiha2004}, not by the Higgs boson of the standard model of particle physics. Putting the constraint $\lambda\lesssim10^{-13}$ together with the bound $\tilde{f}_0\gtrsim 10^2$ inferred from the $\tilde{f}_0-c_s$ parameter space, we found the lower bound $f_0\gtrsim 10^{15}$ in our warm DBI inflation model.

Moreover, we showed that the dissipation parameter in our warm DBI model behaves as $\Upsilon\propto\phi$ which is supported by QFT as well as phenomenological implications. Finally, we examined the swampland criteria in our warm DBI model and concluded that in contrary to the standard inflation which suffers from the violation of the dS limit in the slow-roll regime, the swampland problem can be resolved in our model.

\begin{acknowledgments}
The work of A. Abdolmaleki has been supported financially by Research Institute for Astronomy and Astrophysics of Maragha (RIAAM) under research project No. 1/5237-116.
\end{acknowledgments}










\begin{thebibliography}{105}%
\makeatletter
\providecommand \@ifxundefined [1]{%
 \@ifx{#1\undefined}
}%
\providecommand \@ifnum [1]{%
 \ifnum #1\expandafter \@firstoftwo
 \else \expandafter \@secondoftwo
 \fi
}%
\providecommand \@ifx [1]{%
 \ifx #1\expandafter \@firstoftwo
 \else \expandafter \@secondoftwo
 \fi
}%
\providecommand \natexlab [1]{#1}%
\providecommand \enquote  [1]{``#1''}%
\providecommand \bibnamefont  [1]{#1}%
\providecommand \bibfnamefont [1]{#1}%
\providecommand \citenamefont [1]{#1}%
\providecommand \href@noop [0]{\@secondoftwo}%
\providecommand \href [0]{\begingroup \@sanitize@url \@href}%
\providecommand \@href[1]{\@@startlink{#1}\@@href}%
\providecommand \@@href[1]{\endgroup#1\@@endlink}%
\providecommand \@sanitize@url [0]{\catcode `\\12\catcode `\$12\catcode
  `\&12\catcode `\#12\catcode `\^12\catcode `\_12\catcode `\%12\relax}%
\providecommand \@@startlink[1]{}%
\providecommand \@@endlink[0]{}%
\providecommand \url  [0]{\begingroup\@sanitize@url \@url }%
\providecommand \@url [1]{\endgroup\@href {#1}{\urlprefix }}%
\providecommand \urlprefix  [0]{URL }%
\providecommand \Eprint [0]{\href }%
\providecommand \doibase [0]{http://dx.doi.org/}%
\providecommand \selectlanguage [0]{\@gobble}%
\providecommand \bibinfo  [0]{\@secondoftwo}%
\providecommand \bibfield  [0]{\@secondoftwo}%
\providecommand \translation [1]{[#1]}%
\providecommand \BibitemOpen [0]{}%
\providecommand \bibitemStop [0]{}%
\providecommand \bibitemNoStop [0]{.\EOS\space}%
\providecommand \EOS [0]{\spacefactor3000\relax}%
\providecommand \BibitemShut  [1]{\csname bibitem#1\endcsname}%
\let\auto@bib@innerbib\@empty
\bibitem [{\citenamefont {Starobinsky}(1980)}]{Starobinsky1980}%
  \BibitemOpen
  \bibfield  {author} {\bibinfo {author} {\bibfnamefont {A.}~\bibnamefont
  {Starobinsky}},\ }\href@noop {} {\bibfield  {journal} {\bibinfo  {journal}
  {Phys. Lett. B}\ }\textbf {\bibinfo {volume} {91}},\ \bibinfo {pages} {99}
  (\bibinfo {year} {1980})}\BibitemShut {NoStop}%
\bibitem [{\citenamefont {Sato}(1981{\natexlab{a}})}]{Sato1981}%
  \BibitemOpen
  \bibfield  {author} {\bibinfo {author} {\bibfnamefont {K.}~\bibnamefont
  {Sato}},\ }\href@noop {} {\bibfield  {journal} {\bibinfo  {journal} {Mon.
  Not. R. Astron. Soc.}\ }\textbf {\bibinfo {volume} {195}},\ \bibinfo {pages}
  {467} (\bibinfo {year} {1981}{\natexlab{a}})}\BibitemShut {NoStop}%
\bibitem [{\citenamefont {Sato}(1981{\natexlab{b}})}]{Sato1981-2}%
  \BibitemOpen
  \bibfield  {author} {\bibinfo {author} {\bibfnamefont {K.}~\bibnamefont
  {Sato}},\ }\href@noop {} {\bibfield  {journal} {\bibinfo  {journal} {Phys.
  Lett. B}\ }\textbf {\bibinfo {volume} {99}},\ \bibinfo {pages} {66} (\bibinfo
  {year} {1981}{\natexlab{b}})}\BibitemShut {NoStop}%
\bibitem [{\citenamefont {Guth}(1981)}]{Guth1981}%
  \BibitemOpen
  \bibfield  {author} {\bibinfo {author} {\bibfnamefont {A.~H.}\ \bibnamefont
  {Guth}},\ }\href@noop {} {\bibfield  {journal} {\bibinfo  {journal} {Phys.
  Rev. D}\ }\textbf {\bibinfo {volume} {23}},\ \bibinfo {pages} {347} (\bibinfo
  {year} {1981})}\BibitemShut {NoStop}%
\bibitem [{\citenamefont {Linde}(1982)}]{Linde1982}%
  \BibitemOpen
  \bibfield  {author} {\bibinfo {author} {\bibfnamefont {A.}~\bibnamefont
  {Linde}},\ }\href@noop {} {\bibfield  {journal} {\bibinfo  {journal} {Phys.
  Lett. B}\ }\textbf {\bibinfo {volume} {108}},\ \bibinfo {pages} {389}
  (\bibinfo {year} {1982})}\BibitemShut {NoStop}%
\bibitem [{\citenamefont {Albrecht}\ and\ \citenamefont
  {Steinhardt}(1982)}]{Albrecht1982}%
  \BibitemOpen
  \bibfield  {author} {\bibinfo {author} {\bibfnamefont {A.}~\bibnamefont
  {Albrecht}}\ and\ \bibinfo {author} {\bibfnamefont {P.~J.}\ \bibnamefont
  {Steinhardt}},\ }\href@noop {} {\bibfield  {journal} {\bibinfo  {journal}
  {Phys. Rev. Lett.}\ }\textbf {\bibinfo {volume} {48}},\ \bibinfo {pages}
  {1220} (\bibinfo {year} {1982})}\BibitemShut {NoStop}%
\bibitem [{\citenamefont {Linde}(1983)}]{Linde1983}%
  \BibitemOpen
  \bibfield  {author} {\bibinfo {author} {\bibfnamefont {A.}~\bibnamefont
  {Linde}},\ }\href@noop {} {\bibfield  {journal} {\bibinfo  {journal} {Phys.
  Lett. B}\ }\textbf {\bibinfo {volume} {129}},\ \bibinfo {pages} {177}
  (\bibinfo {year} {1983})}\BibitemShut {NoStop}%
\bibitem [{\citenamefont {Mukhanov}\ and\ \citenamefont
  {Chibisov}(1981)}]{Mukhanov1981}%
  \BibitemOpen
  \bibfield  {author} {\bibinfo {author} {\bibfnamefont {V.}~\bibnamefont
  {Mukhanov}}\ and\ \bibinfo {author} {\bibfnamefont {G.}~\bibnamefont
  {Chibisov}},\ }\href@noop {} {\bibfield  {journal} {\bibinfo  {journal} {JETP
  Lett.}\ }\textbf {\bibinfo {volume} {33}},\ \bibinfo {pages} {532} (\bibinfo
  {year} {1981})}\BibitemShut {NoStop}%
\bibitem [{\citenamefont {Hawking}(1982)}]{Hawking1982}%
  \BibitemOpen
  \bibfield  {author} {\bibinfo {author} {\bibfnamefont {S.}~\bibnamefont
  {Hawking}},\ }\href@noop {} {\bibfield  {journal} {\bibinfo  {journal} {Phys.
  Lett. B}\ }\textbf {\bibinfo {volume} {115}},\ \bibinfo {pages} {295}
  (\bibinfo {year} {1982})}\BibitemShut {NoStop}%
\bibitem [{\citenamefont {Starobinsky}(1982)}]{Starobinsky1982}%
  \BibitemOpen
  \bibfield  {author} {\bibinfo {author} {\bibfnamefont {A.}~\bibnamefont
  {Starobinsky}},\ }\href@noop {} {\bibfield  {journal} {\bibinfo  {journal}
  {Phys. Lett. B}\ }\textbf {\bibinfo {volume} {117}},\ \bibinfo {pages} {175}
  (\bibinfo {year} {1982})}\BibitemShut {NoStop}%
\bibitem [{\citenamefont {Guth}\ and\ \citenamefont {Pi}(1982)}]{Guth1982}%
  \BibitemOpen
  \bibfield  {author} {\bibinfo {author} {\bibfnamefont {A.~H.}\ \bibnamefont
  {Guth}}\ and\ \bibinfo {author} {\bibfnamefont {S.-Y.}\ \bibnamefont {Pi}},\
  }\href@noop {} {\bibfield  {journal} {\bibinfo  {journal} {Phys. Rev. Lett.}\
  }\textbf {\bibinfo {volume} {49}},\ \bibinfo {pages} {1110} (\bibinfo {year}
  {1982})}\BibitemShut {NoStop}%
\bibitem [{\citenamefont {Ade}\ \emph {et~al.}(2016{\natexlab{a}})\citenamefont
  {Ade} \emph {et~al.}}]{Planck2015}%
  \BibitemOpen
  \bibfield  {author} {\bibinfo {author} {\bibfnamefont {P.~A.~R.}\
  \bibnamefont {Ade}} \emph {et~al.},\ }\href@noop {} {\bibfield  {journal}
  {\bibinfo  {journal} {Astron. Astrophys.}\ }\textbf {\bibinfo {volume}
  {594}},\ \bibinfo {pages} {A20} (\bibinfo {year}
  {2016}{\natexlab{a}})}\BibitemShut {NoStop}%
\bibitem [{\citenamefont {Ade}\ \emph {et~al.}(2016{\natexlab{b}})\citenamefont
  {Ade} \emph {et~al.}}]{Planck2015non-Gaussianity}%
  \BibitemOpen
  \bibfield  {author} {\bibinfo {author} {\bibfnamefont {P.~A.~R.}\
  \bibnamefont {Ade}} \emph {et~al.},\ }\href@noop {} {\bibfield  {journal}
  {\bibinfo  {journal} {Astron. Astrophys.}\ }\textbf {\bibinfo {volume}
  {594}},\ \bibinfo {pages} {A17} (\bibinfo {year}
  {2016}{\natexlab{b}})}\BibitemShut {NoStop}%
\bibitem [{\citenamefont {Higgs}(1964)}]{Higgs1964}%
  \BibitemOpen
  \bibfield  {author} {\bibinfo {author} {\bibfnamefont {P.~W.}\ \bibnamefont
  {Higgs}},\ }\href@noop {} {\bibfield  {journal} {\bibinfo  {journal} {Phys.
  Rev. Lett.}\ }\textbf {\bibinfo {volume} {13}},\ \bibinfo {pages} {508}
  (\bibinfo {year} {1964})}\BibitemShut {NoStop}%
\bibitem [{\citenamefont {Aad}\ \emph {et~al.}(2012)\citenamefont {Aad} \emph
  {et~al.}}]{ATLAS2012}%
  \BibitemOpen
  \bibfield  {author} {\bibinfo {author} {\bibfnamefont {G.}~\bibnamefont
  {Aad}} \emph {et~al.},\ }\href@noop {} {\bibfield  {journal} {\bibinfo
  {journal} {Phys. Lett. B}\ }\textbf {\bibinfo {volume} {716}},\ \bibinfo
  {pages} {1} (\bibinfo {year} {2012})}\BibitemShut {NoStop}%
\bibitem [{\citenamefont {Chatrchyan}\ \emph {et~al.}(2012)\citenamefont
  {Chatrchyan} \emph {et~al.}}]{CMS2012}%
  \BibitemOpen
  \bibfield  {author} {\bibinfo {author} {\bibfnamefont {S.}~\bibnamefont
  {Chatrchyan}} \emph {et~al.},\ }\href@noop {} {\bibfield  {journal} {\bibinfo
   {journal} {Phys. Lett. B}\ }\textbf {\bibinfo {volume} {716}},\ \bibinfo
  {pages} {30} (\bibinfo {year} {2012})}\BibitemShut {NoStop}%
\bibitem [{\citenamefont {Pich}()}]{Pich2007}%
  \BibitemOpen
  \bibfield  {author} {\bibinfo {author} {\bibfnamefont {A.}~\bibnamefont
  {Pich}},\ }\href@noop {} {\bibinfo  {journal} {arXiv:0705.4264}\
  }\BibitemShut {NoStop}%
\bibitem [{\citenamefont {Olive}(2016)}]{Patrignani2016}%
  \BibitemOpen
\bibfield  {journal} {  }\bibfield  {author} {\bibinfo {author} {\bibfnamefont
  {K.}~\bibnamefont {Olive}},\ }\href@noop {} {\bibfield  {journal} {\bibinfo
  {journal} {Chinese Phys. C}\ }\textbf {\bibinfo {volume} {40}},\ \bibinfo
  {pages} {100001} (\bibinfo {year} {2016})}\BibitemShut {NoStop}%
\bibitem [{\citenamefont {Germani}\ and\ \citenamefont
  {Kehagias}(2010)}]{Germani2010}%
  \BibitemOpen
  \bibfield  {author} {\bibinfo {author} {\bibfnamefont {C.}~\bibnamefont
  {Germani}}\ and\ \bibinfo {author} {\bibfnamefont {A.}~\bibnamefont
  {Kehagias}},\ }\href@noop {} {\bibfield  {journal} {\bibinfo  {journal}
  {Phys. Rev. Lett.}\ }\textbf {\bibinfo {volume} {105}},\ \bibinfo {pages}
  {011302} (\bibinfo {year} {2010})}\BibitemShut {NoStop}%
\bibitem [{\citenamefont {Copeland}\ \emph {et~al.}(1994)\citenamefont
  {Copeland}, \citenamefont {Liddle}, \citenamefont {Lyth}, \citenamefont
  {Stewart},\ and\ \citenamefont {Wands}}]{Copeland1994}%
  \BibitemOpen
  \bibfield  {author} {\bibinfo {author} {\bibfnamefont {E.~J.}\ \bibnamefont
  {Copeland}}, \bibinfo {author} {\bibfnamefont {A.~R.}\ \bibnamefont
  {Liddle}}, \bibinfo {author} {\bibfnamefont {D.~H.}\ \bibnamefont {Lyth}},
  \bibinfo {author} {\bibfnamefont {E.~D.}\ \bibnamefont {Stewart}}, \ and\
  \bibinfo {author} {\bibfnamefont {D.}~\bibnamefont {Wands}},\ }\href@noop {}
  {\bibfield  {journal} {\bibinfo  {journal} {Phys. Rev. D}\ }\textbf {\bibinfo
  {volume} {49}},\ \bibinfo {pages} {6410} (\bibinfo {year}
  {1994})}\BibitemShut {NoStop}%
\bibitem [{\citenamefont {Cervantes-Cota}\ and\ \citenamefont
  {Dehnen}(1995)}]{Cervantes-Cota1995}%
  \BibitemOpen
  \bibfield  {author} {\bibinfo {author} {\bibfnamefont {J.}~\bibnamefont
  {Cervantes-Cota}}\ and\ \bibinfo {author} {\bibfnamefont {H.}~\bibnamefont
  {Dehnen}},\ }\href@noop {} {\bibfield  {journal} {\bibinfo  {journal} {Nucl.
  Phys. B}\ }\textbf {\bibinfo {volume} {442}},\ \bibinfo {pages} {391}
  (\bibinfo {year} {1995})}\BibitemShut {NoStop}%
\bibitem [{\citenamefont {Bezrukov}\ and\ \citenamefont
  {Shaposhnikov}(2008)}]{Bezrukov2008}%
  \BibitemOpen
  \bibfield  {author} {\bibinfo {author} {\bibfnamefont {F.}~\bibnamefont
  {Bezrukov}}\ and\ \bibinfo {author} {\bibfnamefont {M.}~\bibnamefont
  {Shaposhnikov}},\ }\href@noop {} {\bibfield  {journal} {\bibinfo  {journal}
  {Phys. Lett. B}\ }\textbf {\bibinfo {volume} {659}},\ \bibinfo {pages} {703}
  (\bibinfo {year} {2008})}\BibitemShut {NoStop}%
\bibitem [{\citenamefont {Escriv{\`a}}\ and\ \citenamefont
  {Germani}(2017)}]{Germani2017}%
  \BibitemOpen
  \bibfield  {author} {\bibinfo {author} {\bibfnamefont {A.}~\bibnamefont
  {Escriv{\`a}}}\ and\ \bibinfo {author} {\bibfnamefont {C.}~\bibnamefont
  {Germani}},\ }\href@noop {} {\bibfield  {journal} {\bibinfo  {journal} {Phys.
  Rev. D}\ }\textbf {\bibinfo {volume} {95}},\ \bibinfo {pages} {123526}
  (\bibinfo {year} {2017})}\BibitemShut {NoStop}%
\bibitem [{\citenamefont {Calmet}\ \emph {et~al.}(2017)\citenamefont {Calmet},
  \citenamefont {Kuntz},\ and\ \citenamefont {Moss}}]{Calmet2017}%
  \BibitemOpen
  \bibfield  {author} {\bibinfo {author} {\bibfnamefont {X.}~\bibnamefont
  {Calmet}}, \bibinfo {author} {\bibfnamefont {I.}~\bibnamefont {Kuntz}}, \
  and\ \bibinfo {author} {\bibfnamefont {I.~G.}\ \bibnamefont {Moss}},\ }\href
  {http://arxiv.org/abs/1701.02140} {\  (\bibinfo {year} {2017})},\ \Eprint
  {http://arxiv.org/abs/1701.02140} {arXiv:1701.02140} \BibitemShut {NoStop}%
\bibitem [{\citenamefont {Ballesteros}\ \emph {et~al.}(2017)\citenamefont
  {Ballesteros}, \citenamefont {Redondo}, \citenamefont {Ringwald},\ and\
  \citenamefont {Tamarit}}]{Ballesteros2017}%
  \BibitemOpen
  \bibfield  {author} {\bibinfo {author} {\bibfnamefont {G.}~\bibnamefont
  {Ballesteros}}, \bibinfo {author} {\bibfnamefont {J.}~\bibnamefont
  {Redondo}}, \bibinfo {author} {\bibfnamefont {A.}~\bibnamefont {Ringwald}}, \
  and\ \bibinfo {author} {\bibfnamefont {C.}~\bibnamefont {Tamarit}},\
  }\href@noop {} {\bibfield  {journal} {\bibinfo  {journal} {J. Cosmol.
  Astropart. Phys.}\ }\textbf {\bibinfo {volume} {08}},\ \bibinfo {pages} {001}
  (\bibinfo {year} {2017})}\BibitemShut {NoStop}%
\bibitem [{\citenamefont {Agrawal}\ \emph {et~al.}(2018)\citenamefont
  {Agrawal}, \citenamefont {Obied}, \citenamefont {Steinhardt},\ and\
  \citenamefont {Vafa}}]{Agrawal2018}%
  \BibitemOpen
  \bibfield  {author} {\bibinfo {author} {\bibfnamefont {P.}~\bibnamefont
  {Agrawal}}, \bibinfo {author} {\bibfnamefont {G.}~\bibnamefont {Obied}},
  \bibinfo {author} {\bibfnamefont {P.~J.}\ \bibnamefont {Steinhardt}}, \ and\
  \bibinfo {author} {\bibfnamefont {C.}~\bibnamefont {Vafa}},\ }\href@noop {}
  {\bibfield  {journal} {\bibinfo  {journal} {Physics Letters B}\ }\textbf
  {\bibinfo {volume} {784}},\ \bibinfo {pages} {271} (\bibinfo {year}
  {2018})}\BibitemShut {NoStop}%
\bibitem [{\citenamefont {Ooguri}\ \emph {et~al.}(2018)\citenamefont {Ooguri},
  \citenamefont {Palti}, \citenamefont {Shiu},\ and\ \citenamefont
  {Vafa}}]{Ooguri2018}%
  \BibitemOpen
  \bibfield  {author} {\bibinfo {author} {\bibfnamefont {H.}~\bibnamefont
  {Ooguri}}, \bibinfo {author} {\bibfnamefont {E.}~\bibnamefont {Palti}},
  \bibinfo {author} {\bibfnamefont {G.}~\bibnamefont {Shiu}}, \ and\ \bibinfo
  {author} {\bibfnamefont {C.}~\bibnamefont {Vafa}},\ }\href@noop {} {\bibfield
   {journal} {\bibinfo  {journal} {Physics Letters B}\ } (\bibinfo {year}
  {2018})}\BibitemShut {NoStop}%
\bibitem [{\citenamefont {Silverstein}\ and\ \citenamefont
  {Tong}(2004)}]{Silverstein2004}%
  \BibitemOpen
  \bibfield  {author} {\bibinfo {author} {\bibfnamefont {E.}~\bibnamefont
  {Silverstein}}\ and\ \bibinfo {author} {\bibfnamefont {D.}~\bibnamefont
  {Tong}},\ }\href@noop {} {\bibfield  {journal} {\bibinfo  {journal} {Phys.
  Rev. D}\ }\textbf {\bibinfo {volume} {70}},\ \bibinfo {pages} {103505}
  (\bibinfo {year} {2004})}\BibitemShut {NoStop}%
\bibitem [{\citenamefont {Alishahiha}\ \emph {et~al.}(2004)\citenamefont
  {Alishahiha}, \citenamefont {Silverstein},\ and\ \citenamefont
  {Tong}}]{Alishahiha2004}%
  \BibitemOpen
  \bibfield  {author} {\bibinfo {author} {\bibfnamefont {M.}~\bibnamefont
  {Alishahiha}}, \bibinfo {author} {\bibfnamefont {E.}~\bibnamefont
  {Silverstein}}, \ and\ \bibinfo {author} {\bibfnamefont {D.}~\bibnamefont
  {Tong}},\ }\href@noop {} {\bibfield  {journal} {\bibinfo  {journal} {Phys.
  Rev. D}\ }\textbf {\bibinfo {volume} {70}},\ \bibinfo {pages} {123505}
  (\bibinfo {year} {2004})}\BibitemShut {NoStop}%
\bibitem [{\citenamefont {Chen}(2005{\natexlab{a}})}]{Chen2005}%
  \BibitemOpen
  \bibfield  {author} {\bibinfo {author} {\bibfnamefont {X.}~\bibnamefont
  {Chen}},\ }\href@noop {} {\bibfield  {journal} {\bibinfo  {journal} {J. High
  Energy Phys.}\ }\textbf {\bibinfo {volume} {08}},\ \bibinfo {pages} {045}
  (\bibinfo {year} {2005}{\natexlab{a}})}\BibitemShut {NoStop}%
\bibitem [{\citenamefont {Chen}(2005{\natexlab{b}})}]{Chen2005-2}%
  \BibitemOpen
  \bibfield  {author} {\bibinfo {author} {\bibfnamefont {X.}~\bibnamefont
  {Chen}},\ }\href@noop {} {\bibfield  {journal} {\bibinfo  {journal} {Phys.
  Rev. D}\ }\textbf {\bibinfo {volume} {71}},\ \bibinfo {pages} {063506}
  (\bibinfo {year} {2005}{\natexlab{b}})}\BibitemShut {NoStop}%
\bibitem [{\citenamefont {Armend{\'a}riz-Pic{\'o}n}\ \emph
  {et~al.}(1999)\citenamefont {Armend{\'a}riz-Pic{\'o}n}, \citenamefont
  {Damour},\ and\ \citenamefont {Mukhanov}}]{Armendariz-Picon1999}%
  \BibitemOpen
  \bibfield  {author} {\bibinfo {author} {\bibfnamefont {C.}~\bibnamefont
  {Armend{\'a}riz-Pic{\'o}n}}, \bibinfo {author} {\bibfnamefont
  {T.}~\bibnamefont {Damour}}, \ and\ \bibinfo {author} {\bibfnamefont
  {V.}~\bibnamefont {Mukhanov}},\ }\href@noop {} {\bibfield  {journal}
  {\bibinfo  {journal} {Phys. Lett. B}\ }\textbf {\bibinfo {volume} {458}},\
  \bibinfo {pages} {209} (\bibinfo {year} {1999})}\BibitemShut {NoStop}%
\bibitem [{\citenamefont {Garriga}\ and\ \citenamefont
  {Mukhanov}(1999)}]{Garriga1999}%
  \BibitemOpen
  \bibfield  {author} {\bibinfo {author} {\bibfnamefont {J.}~\bibnamefont
  {Garriga}}\ and\ \bibinfo {author} {\bibfnamefont {V.}~\bibnamefont
  {Mukhanov}},\ }\href@noop {} {\bibfield  {journal} {\bibinfo  {journal}
  {Phys. Lett. B}\ }\textbf {\bibinfo {volume} {458}},\ \bibinfo {pages} {219}
  (\bibinfo {year} {1999})}\BibitemShut {NoStop}%
\bibitem [{\citenamefont {Mukhanov}\ and\ \citenamefont
  {Vikman}(2006)}]{Mukhanov2006}%
  \BibitemOpen
  \bibfield  {author} {\bibinfo {author} {\bibfnamefont {V.}~\bibnamefont
  {Mukhanov}}\ and\ \bibinfo {author} {\bibfnamefont {A.}~\bibnamefont
  {Vikman}},\ }\href@noop {} {\bibfield  {journal} {\bibinfo  {journal} {J.
  Cosmol. Astropart. Phys.}\ }\textbf {\bibinfo {volume} {02}},\ \bibinfo
  {pages} {004} (\bibinfo {year} {2006})}\BibitemShut {NoStop}%
\bibitem [{\citenamefont {Helmer}\ and\ \citenamefont
  {Winitzki}(2006)}]{Helmer2006}%
  \BibitemOpen
  \bibfield  {author} {\bibinfo {author} {\bibfnamefont {F.}~\bibnamefont
  {Helmer}}\ and\ \bibinfo {author} {\bibfnamefont {S.}~\bibnamefont
  {Winitzki}},\ }\href@noop {} {\bibfield  {journal} {\bibinfo  {journal}
  {Phys. Rev. D}\ }\textbf {\bibinfo {volume} {74}},\ \bibinfo {pages} {063528}
  (\bibinfo {year} {2006})}\BibitemShut {NoStop}%
\bibitem [{\citenamefont {Taveras}\ and\ \citenamefont
  {Yunes}(2008)}]{Taveras2008}%
  \BibitemOpen
  \bibfield  {author} {\bibinfo {author} {\bibfnamefont {V.}~\bibnamefont
  {Taveras}}\ and\ \bibinfo {author} {\bibfnamefont {N.}~\bibnamefont
  {Yunes}},\ }\href@noop {} {\bibfield  {journal} {\bibinfo  {journal} {Phys.
  Rev. D}\ }\textbf {\bibinfo {volume} {78}},\ \bibinfo {pages} {064070}
  (\bibinfo {year} {2008})}\BibitemShut {NoStop}%
\bibitem [{\citenamefont {Bose}\ and\ \citenamefont
  {Majumdar}(2009{\natexlab{a}})}]{Bose2009}%
  \BibitemOpen
  \bibfield  {author} {\bibinfo {author} {\bibfnamefont {N.}~\bibnamefont
  {Bose}}\ and\ \bibinfo {author} {\bibfnamefont {a.~S.}\ \bibnamefont
  {Majumdar}},\ }\href@noop {} {\bibfield  {journal} {\bibinfo  {journal}
  {Phys. Rev. D}\ }\textbf {\bibinfo {volume} {80}},\ \bibinfo {pages} {103508}
  (\bibinfo {year} {2009}{\natexlab{a}})}\BibitemShut {NoStop}%
\bibitem [{\citenamefont {Bose}\ and\ \citenamefont
  {Majumdar}(2009{\natexlab{b}})}]{Bose2009-2}%
  \BibitemOpen
  \bibfield  {author} {\bibinfo {author} {\bibfnamefont {N.}~\bibnamefont
  {Bose}}\ and\ \bibinfo {author} {\bibfnamefont {A.~S.}\ \bibnamefont
  {Majumdar}},\ }\href@noop {} {\bibfield  {journal} {\bibinfo  {journal}
  {Phys. Rev. D}\ }\textbf {\bibinfo {volume} {79}},\ \bibinfo {pages} {103517}
  (\bibinfo {year} {2009}{\natexlab{b}})}\BibitemShut {NoStop}%
\bibitem [{\citenamefont {Franche}\ \emph
  {et~al.}(2010{\natexlab{a}})\citenamefont {Franche}, \citenamefont {Gwyn},
  \citenamefont {Underwood},\ and\ \citenamefont {Wissanji}}]{Franche2010}%
  \BibitemOpen
  \bibfield  {author} {\bibinfo {author} {\bibfnamefont {P.}~\bibnamefont
  {Franche}}, \bibinfo {author} {\bibfnamefont {R.}~\bibnamefont {Gwyn}},
  \bibinfo {author} {\bibfnamefont {B.}~\bibnamefont {Underwood}}, \ and\
  \bibinfo {author} {\bibfnamefont {A.}~\bibnamefont {Wissanji}},\ }\href@noop
  {} {\bibfield  {journal} {\bibinfo  {journal} {Phys. Rev. D}\ }\textbf
  {\bibinfo {volume} {81}},\ \bibinfo {pages} {123526} (\bibinfo {year}
  {2010}{\natexlab{a}})}\BibitemShut {NoStop}%
\bibitem [{\citenamefont {Franche}\ \emph
  {et~al.}(2010{\natexlab{b}})\citenamefont {Franche}, \citenamefont {Gwyn},
  \citenamefont {Underwood},\ and\ \citenamefont {Wissanji}}]{Franche2010-2}%
  \BibitemOpen
  \bibfield  {author} {\bibinfo {author} {\bibfnamefont {P.}~\bibnamefont
  {Franche}}, \bibinfo {author} {\bibfnamefont {R.}~\bibnamefont {Gwyn}},
  \bibinfo {author} {\bibfnamefont {B.}~\bibnamefont {Underwood}}, \ and\
  \bibinfo {author} {\bibfnamefont {A.}~\bibnamefont {Wissanji}},\ }\href@noop
  {} {\bibfield  {journal} {\bibinfo  {journal} {Phys. Rev. D}\ }\textbf
  {\bibinfo {volume} {82}},\ \bibinfo {pages} {063528} (\bibinfo {year}
  {2010}{\natexlab{b}})}\BibitemShut {NoStop}%
\bibitem [{\citenamefont {Devi}\ \emph {et~al.}(2011)\citenamefont {Devi},
  \citenamefont {Nautiyal},\ and\ \citenamefont {Sen}}]{Devi2011}%
  \BibitemOpen
  \bibfield  {author} {\bibinfo {author} {\bibfnamefont {N.~C.}\ \bibnamefont
  {Devi}}, \bibinfo {author} {\bibfnamefont {A.}~\bibnamefont {Nautiyal}}, \
  and\ \bibinfo {author} {\bibfnamefont {A.~A.}\ \bibnamefont {Sen}},\
  }\href@noop {} {\bibfield  {journal} {\bibinfo  {journal} {Phys. Rev. D}\
  }\textbf {\bibinfo {volume} {84}},\ \bibinfo {pages} {103504} (\bibinfo
  {year} {2011})}\BibitemShut {NoStop}%
\bibitem [{\citenamefont {Unnikrishnan}\ \emph {et~al.}(2012)\citenamefont
  {Unnikrishnan}, \citenamefont {Sahni},\ and\ \citenamefont
  {Toporensky}}]{Unnikrishnan2012}%
  \BibitemOpen
  \bibfield  {author} {\bibinfo {author} {\bibfnamefont {S.}~\bibnamefont
  {Unnikrishnan}}, \bibinfo {author} {\bibfnamefont {V.}~\bibnamefont {Sahni}},
  \ and\ \bibinfo {author} {\bibfnamefont {A.}~\bibnamefont {Toporensky}},\
  }\href@noop {} {\bibfield  {journal} {\bibinfo  {journal} {J. Cosmol.
  Astropart. Phys.}\ }\textbf {\bibinfo {volume} {08}},\ \bibinfo {pages} {018}
  (\bibinfo {year} {2012})}\BibitemShut {NoStop}%
\bibitem [{\citenamefont {Rezazadeh}\ \emph {et~al.}(2015)\citenamefont
  {Rezazadeh}, \citenamefont {Karami},\ and\ \citenamefont
  {Karimi}}]{Rezazadeh2015}%
  \BibitemOpen
  \bibfield  {author} {\bibinfo {author} {\bibfnamefont {K.}~\bibnamefont
  {Rezazadeh}}, \bibinfo {author} {\bibfnamefont {K.}~\bibnamefont {Karami}}, \
  and\ \bibinfo {author} {\bibfnamefont {P.}~\bibnamefont {Karimi}},\
  }\href@noop {} {\bibfield  {journal} {\bibinfo  {journal} {J. Cosmol.
  Astropart. Phys.}\ }\textbf {\bibinfo {volume} {09}},\ \bibinfo {pages} {053}
  (\bibinfo {year} {2015})}\BibitemShut {NoStop}%
\bibitem [{\citenamefont {Rezazadeh}\ \emph {et~al.}(2017)\citenamefont
  {Rezazadeh}, \citenamefont {Karami},\ and\ \citenamefont
  {Hashemi}}]{Rezazadeh2017-2}%
  \BibitemOpen
  \bibfield  {author} {\bibinfo {author} {\bibfnamefont {K.}~\bibnamefont
  {Rezazadeh}}, \bibinfo {author} {\bibfnamefont {K.}~\bibnamefont {Karami}}, \
  and\ \bibinfo {author} {\bibfnamefont {S.}~\bibnamefont {Hashemi}},\
  }\href@noop {} {\bibfield  {journal} {\bibinfo  {journal} {Phys. Rev. D}\
  }\textbf {\bibinfo {volume} {95}},\ \bibinfo {pages} {103506} (\bibinfo
  {year} {2017})}\BibitemShut {NoStop}%
\bibitem [{\citenamefont {Chen}\ \emph {et~al.}(2007)\citenamefont {Chen},
  \citenamefont {Huang}, \citenamefont {Kachru},\ and\ \citenamefont
  {Shiu}}]{Chen2007}%
  \BibitemOpen
  \bibfield  {author} {\bibinfo {author} {\bibfnamefont {X.}~\bibnamefont
  {Chen}}, \bibinfo {author} {\bibfnamefont {M.-x.}\ \bibnamefont {Huang}},
  \bibinfo {author} {\bibfnamefont {S.}~\bibnamefont {Kachru}}, \ and\ \bibinfo
  {author} {\bibfnamefont {G.}~\bibnamefont {Shiu}},\ }\href@noop {} {\bibfield
   {journal} {\bibinfo  {journal} {J. Cosmol. Astropart. Phys.}\ }\textbf
  {\bibinfo {volume} {01}},\ \bibinfo {pages} {002} (\bibinfo {year}
  {2007})}\BibitemShut {NoStop}%
\bibitem [{\citenamefont {Li}\ \emph {et~al.}(2008)\citenamefont {Li},
  \citenamefont {Wang},\ and\ \citenamefont {Wang}}]{Li2008}%
  \BibitemOpen
  \bibfield  {author} {\bibinfo {author} {\bibfnamefont {M.}~\bibnamefont
  {Li}}, \bibinfo {author} {\bibfnamefont {T.}~\bibnamefont {Wang}}, \ and\
  \bibinfo {author} {\bibfnamefont {Y.}~\bibnamefont {Wang}},\ }\href@noop {}
  {\bibfield  {journal} {\bibinfo  {journal} {J. Cosmol. Astropart. Phys.}\
  }\textbf {\bibinfo {volume} {03}},\ \bibinfo {pages} {028} (\bibinfo {year}
  {2008})}\BibitemShut {NoStop}%
\bibitem [{\citenamefont {Tolley}\ and\ \citenamefont
  {Wyman}(2010)}]{Tolley2010}%
  \BibitemOpen
  \bibfield  {author} {\bibinfo {author} {\bibfnamefont {A.~J.}\ \bibnamefont
  {Tolley}}\ and\ \bibinfo {author} {\bibfnamefont {M.}~\bibnamefont {Wyman}},\
  }\href@noop {} {\bibfield  {journal} {\bibinfo  {journal} {Phys. Rev. D}\
  }\textbf {\bibinfo {volume} {81}},\ \bibinfo {pages} {043502} (\bibinfo
  {year} {2010})}\BibitemShut {NoStop}%
\bibitem [{\citenamefont {Seery}\ and\ \citenamefont
  {Lidsey}(2005)}]{Seery2005}%
  \BibitemOpen
  \bibfield  {author} {\bibinfo {author} {\bibfnamefont {D.}~\bibnamefont
  {Seery}}\ and\ \bibinfo {author} {\bibfnamefont {J.~E.}\ \bibnamefont
  {Lidsey}},\ }\href@noop {} {\bibfield  {journal} {\bibinfo  {journal} {J.
  Cosmol. Astropart. Phys.}\ }\textbf {\bibinfo {volume} {06}},\ \bibinfo
  {pages} {003} (\bibinfo {year} {2005})}\BibitemShut {NoStop}%
\bibitem [{\citenamefont {Panotopoulos}(2007)}]{Panotopoulos2007}%
  \BibitemOpen
  \bibfield  {author} {\bibinfo {author} {\bibfnamefont {G.}~\bibnamefont
  {Panotopoulos}},\ }\href@noop {} {\bibfield  {journal} {\bibinfo  {journal}
  {Phys. Rev. D}\ }\textbf {\bibinfo {volume} {76}},\ \bibinfo {pages} {127302}
  (\bibinfo {year} {2007})}\BibitemShut {NoStop}%
\bibitem [{\citenamefont {Maldacena}(2003)}]{Maldacena2003}%
  \BibitemOpen
  \bibfield  {author} {\bibinfo {author} {\bibfnamefont {J.}~\bibnamefont
  {Maldacena}},\ }\href@noop {} {\bibfield  {journal} {\bibinfo  {journal} {J.
  High Energy Phys.}\ }\textbf {\bibinfo {volume} {05}},\ \bibinfo {pages}
  {013} (\bibinfo {year} {2003})}\BibitemShut {NoStop}%
\bibitem [{\citenamefont {Acquaviva}\ \emph {et~al.}(2003)\citenamefont
  {Acquaviva}, \citenamefont {Bartolo}, \citenamefont {Matarrese},\ and\
  \citenamefont {Riotto}}]{Acquaviva2003}%
  \BibitemOpen
  \bibfield  {author} {\bibinfo {author} {\bibfnamefont {V.}~\bibnamefont
  {Acquaviva}}, \bibinfo {author} {\bibfnamefont {N.}~\bibnamefont {Bartolo}},
  \bibinfo {author} {\bibfnamefont {S.}~\bibnamefont {Matarrese}}, \ and\
  \bibinfo {author} {\bibfnamefont {A.}~\bibnamefont {Riotto}},\ }\href@noop {}
  {\bibfield  {journal} {\bibinfo  {journal} {Nucl. Phys. B}\ }\textbf
  {\bibinfo {volume} {667}},\ \bibinfo {pages} {119} (\bibinfo {year}
  {2003})}\BibitemShut {NoStop}%
\bibitem [{\citenamefont {Rigopoulos}\ \emph {et~al.}(2005)\citenamefont
  {Rigopoulos}, \citenamefont {Shellard},\ and\ \citenamefont {van
  Tent}}]{Rigopoulos2005}%
  \BibitemOpen
  \bibfield  {author} {\bibinfo {author} {\bibfnamefont {G.~I.}\ \bibnamefont
  {Rigopoulos}}, \bibinfo {author} {\bibfnamefont {E.~P.~S.}\ \bibnamefont
  {Shellard}}, \ and\ \bibinfo {author} {\bibfnamefont {B.~J.~W.}\ \bibnamefont
  {van Tent}},\ }\href@noop {} {\bibfield  {journal} {\bibinfo  {journal}
  {Phys. Rev. D}\ }\textbf {\bibinfo {volume} {72}},\ \bibinfo {pages} {083507}
  (\bibinfo {year} {2005})}\BibitemShut {NoStop}%
\bibitem [{\citenamefont {Creminelli}\ and\ \citenamefont
  {Zaldarriaga}(2004)}]{Creminelli2004}%
  \BibitemOpen
  \bibfield  {author} {\bibinfo {author} {\bibfnamefont {P.}~\bibnamefont
  {Creminelli}}\ and\ \bibinfo {author} {\bibfnamefont {M.}~\bibnamefont
  {Zaldarriaga}},\ }\href@noop {} {\bibfield  {journal} {\bibinfo  {journal}
  {J. Cosmol. Astropart. Phys.}\ }\textbf {\bibinfo {volume} {10}},\ \bibinfo
  {pages} {006} (\bibinfo {year} {2004})}\BibitemShut {NoStop}%
\bibitem [{\citenamefont {Bassett}\ \emph {et~al.}(2006)\citenamefont
  {Bassett}, \citenamefont {Tsujikawa},\ and\ \citenamefont
  {Wands}}]{Basset2006}%
  \BibitemOpen
  \bibfield  {author} {\bibinfo {author} {\bibfnamefont {B.~A.}\ \bibnamefont
  {Bassett}}, \bibinfo {author} {\bibfnamefont {S.}~\bibnamefont {Tsujikawa}},
  \ and\ \bibinfo {author} {\bibfnamefont {D.}~\bibnamefont {Wands}},\
  }\href@noop {} {\bibfield  {journal} {\bibinfo  {journal} {Rev. Mod. Phys.}\
  }\textbf {\bibinfo {volume} {78}},\ \bibinfo {pages} {537} (\bibinfo {year}
  {2006})}\BibitemShut {NoStop}%
\bibitem [{\citenamefont {Berera}\ and\ \citenamefont
  {Fang}(1995)}]{Berera1995}%
  \BibitemOpen
  \bibfield  {author} {\bibinfo {author} {\bibfnamefont {A.}~\bibnamefont
  {Berera}}\ and\ \bibinfo {author} {\bibfnamefont {L.-Z.}\ \bibnamefont
  {Fang}},\ }\href@noop {} {\bibfield  {journal} {\bibinfo  {journal} {Phys.
  Rev. Lett.}\ }\textbf {\bibinfo {volume} {74}},\ \bibinfo {pages} {1912}
  (\bibinfo {year} {1995})}\BibitemShut {NoStop}%
\bibitem [{\citenamefont {Berera}(1995)}]{Berera1995-2}%
  \BibitemOpen
  \bibfield  {author} {\bibinfo {author} {\bibfnamefont {A.}~\bibnamefont
  {Berera}},\ }\href@noop {} {\bibfield  {journal} {\bibinfo  {journal} {Phys.
  Rev. Lett.}\ }\textbf {\bibinfo {volume} {75}},\ \bibinfo {pages} {3218}
  (\bibinfo {year} {1995})}\BibitemShut {NoStop}%
\bibitem [{\citenamefont {Berera}(1997)}]{Berera1997}%
  \BibitemOpen
  \bibfield  {author} {\bibinfo {author} {\bibfnamefont {A.}~\bibnamefont
  {Berera}},\ }\href@noop {} {\bibfield  {journal} {\bibinfo  {journal} {Phys.
  Rev. D}\ }\textbf {\bibinfo {volume} {55}},\ \bibinfo {pages} {3346}
  (\bibinfo {year} {1997})}\BibitemShut {NoStop}%
\bibitem [{\citenamefont {Berera}(2000)}]{Berera2000}%
  \BibitemOpen
  \bibfield  {author} {\bibinfo {author} {\bibfnamefont {A.}~\bibnamefont
  {Berera}},\ }\href@noop {} {\bibfield  {journal} {\bibinfo  {journal} {Nucl.
  Phys. B}\ }\textbf {\bibinfo {volume} {585}},\ \bibinfo {pages} {666}
  (\bibinfo {year} {2000})}\BibitemShut {NoStop}%
\bibitem [{\citenamefont {Taylor}\ and\ \citenamefont
  {Berera}(2000)}]{Taylor2000}%
  \BibitemOpen
  \bibfield  {author} {\bibinfo {author} {\bibfnamefont {A.~N.}\ \bibnamefont
  {Taylor}}\ and\ \bibinfo {author} {\bibfnamefont {A.}~\bibnamefont
  {Berera}},\ }\href@noop {} {\bibfield  {journal} {\bibinfo  {journal} {Phys.
  Rev. D}\ }\textbf {\bibinfo {volume} {62}},\ \bibinfo {pages} {083517}
  (\bibinfo {year} {2000})}\BibitemShut {NoStop}%
\bibitem [{\citenamefont {Hall}\ \emph {et~al.}(2004)\citenamefont {Hall},
  \citenamefont {Moss},\ and\ \citenamefont {Berera}}]{Hall2004}%
  \BibitemOpen
  \bibfield  {author} {\bibinfo {author} {\bibfnamefont {L.~M.~H.}\
  \bibnamefont {Hall}}, \bibinfo {author} {\bibfnamefont {I.~G.}\ \bibnamefont
  {Moss}}, \ and\ \bibinfo {author} {\bibfnamefont {A.}~\bibnamefont
  {Berera}},\ }\href@noop {} {\bibfield  {journal} {\bibinfo  {journal} {Phys.
  Rev. D}\ }\textbf {\bibinfo {volume} {69}},\ \bibinfo {pages} {083525}
  (\bibinfo {year} {2004})}\BibitemShut {NoStop}%
\bibitem [{\citenamefont {Moss}\ and\ \citenamefont {Xiong}(2007)}]{Moss2007}%
  \BibitemOpen
  \bibfield  {author} {\bibinfo {author} {\bibfnamefont {I.~G.}\ \bibnamefont
  {Moss}}\ and\ \bibinfo {author} {\bibfnamefont {C.}~\bibnamefont {Xiong}},\
  }\href@noop {} {\bibfield  {journal} {\bibinfo  {journal} {J. Cosmol.
  Astropart. Phys.}\ }\textbf {\bibinfo {volume} {04}},\ \bibinfo {pages} {007}
  (\bibinfo {year} {2007})}\BibitemShut {NoStop}%
\bibitem [{\citenamefont {Graham}\ and\ \citenamefont
  {Moss}(2009)}]{Graham2009}%
  \BibitemOpen
  \bibfield  {author} {\bibinfo {author} {\bibfnamefont {C.}~\bibnamefont
  {Graham}}\ and\ \bibinfo {author} {\bibfnamefont {I.~G.}\ \bibnamefont
  {Moss}},\ }\href@noop {} {\bibfield  {journal} {\bibinfo  {journal} {J.
  Cosmol. Astropart. Phys.}\ }\textbf {\bibinfo {volume} {07}},\ \bibinfo
  {pages} {013} (\bibinfo {year} {2009})}\BibitemShut {NoStop}%
\bibitem [{\citenamefont {Ramos}\ and\ \citenamefont
  {da~Silva}(2013)}]{Ramos2013}%
  \BibitemOpen
  \bibfield  {author} {\bibinfo {author} {\bibfnamefont {R.~O.}\ \bibnamefont
  {Ramos}}\ and\ \bibinfo {author} {\bibfnamefont {L.}~\bibnamefont
  {da~Silva}},\ }\href@noop {} {\bibfield  {journal} {\bibinfo  {journal} {J.
  Cosmol. Astropart. Phys.}\ }\textbf {\bibinfo {volume} {03}},\ \bibinfo
  {pages} {032} (\bibinfo {year} {2013})}\BibitemShut {NoStop}%
\bibitem [{\citenamefont {Bartrum}\ \emph {et~al.}(2014)\citenamefont
  {Bartrum}, \citenamefont {Bastero-Gil}, \citenamefont {Berera}, \citenamefont
  {Cerezo}, \citenamefont {Ramos},\ and\ \citenamefont {Rosa}}]{Bartrum2014}%
  \BibitemOpen
  \bibfield  {author} {\bibinfo {author} {\bibfnamefont {S.}~\bibnamefont
  {Bartrum}}, \bibinfo {author} {\bibfnamefont {M.}~\bibnamefont
  {Bastero-Gil}}, \bibinfo {author} {\bibfnamefont {A.}~\bibnamefont {Berera}},
  \bibinfo {author} {\bibfnamefont {R.}~\bibnamefont {Cerezo}}, \bibinfo
  {author} {\bibfnamefont {R.~O.}\ \bibnamefont {Ramos}}, \ and\ \bibinfo
  {author} {\bibfnamefont {J.~G.}\ \bibnamefont {Rosa}},\ }\href@noop {}
  {\bibfield  {journal} {\bibinfo  {journal} {Phys. Lett. B}\ }\textbf
  {\bibinfo {volume} {732}},\ \bibinfo {pages} {116} (\bibinfo {year}
  {2014})}\BibitemShut {NoStop}%
\bibitem [{\citenamefont {Bastero-Gil}\ \emph {et~al.}(2014)\citenamefont
  {Bastero-Gil}, \citenamefont {Berera}, \citenamefont {Moss},\ and\
  \citenamefont {Ramos}}]{Bastero-Gil2014}%
  \BibitemOpen
  \bibfield  {author} {\bibinfo {author} {\bibfnamefont {M.}~\bibnamefont
  {Bastero-Gil}}, \bibinfo {author} {\bibfnamefont {A.}~\bibnamefont {Berera}},
  \bibinfo {author} {\bibfnamefont {I.~G.}\ \bibnamefont {Moss}}, \ and\
  \bibinfo {author} {\bibfnamefont {R.~O.}\ \bibnamefont {Ramos}},\ }\href@noop
  {} {\bibfield  {journal} {\bibinfo  {journal} {J. Cosmol. Astropart. Phys.}\
  }\textbf {\bibinfo {volume} {05}},\ \bibinfo {pages} {004} (\bibinfo {year}
  {2014})}\BibitemShut {NoStop}%
\bibitem [{\citenamefont {Bastero-Gil}\ \emph {et~al.}(2016)\citenamefont
  {Bastero-Gil}, \citenamefont {Berera}, \citenamefont {Ramos},\ and\
  \citenamefont {Rosa}}]{Bastero-Gil2016}%
  \BibitemOpen
  \bibfield  {author} {\bibinfo {author} {\bibfnamefont {M.}~\bibnamefont
  {Bastero-Gil}}, \bibinfo {author} {\bibfnamefont {A.}~\bibnamefont {Berera}},
  \bibinfo {author} {\bibfnamefont {R.~O.}\ \bibnamefont {Ramos}}, \ and\
  \bibinfo {author} {\bibfnamefont {J.~G.}\ \bibnamefont {Rosa}},\ }\href@noop
  {} {\bibfield  {journal} {\bibinfo  {journal} {Phys. Rev. Lett.}\ }\textbf
  {\bibinfo {volume} {117}},\ \bibinfo {pages} {151301} (\bibinfo {year}
  {2016})}\BibitemShut {NoStop}%
\bibitem [{\citenamefont {Benetti}\ and\ \citenamefont
  {Ramos}(2017)}]{Benetti2017}%
  \BibitemOpen
  \bibfield  {author} {\bibinfo {author} {\bibfnamefont {M.}~\bibnamefont
  {Benetti}}\ and\ \bibinfo {author} {\bibfnamefont {R.~O.}\ \bibnamefont
  {Ramos}},\ }\href@noop {} {\bibfield  {journal} {\bibinfo  {journal} {Phys.
  Rev. D}\ }\textbf {\bibinfo {volume} {95}},\ \bibinfo {pages} {023517}
  (\bibinfo {year} {2017})}\BibitemShut {NoStop}%
\bibitem [{\citenamefont {Bastero-Gil}\ and\ \citenamefont
  {Berera}(2009)}]{Bastero-Gil2009}%
  \BibitemOpen
  \bibfield  {author} {\bibinfo {author} {\bibfnamefont {M.}~\bibnamefont
  {Bastero-Gil}}\ and\ \bibinfo {author} {\bibfnamefont {A.}~\bibnamefont
  {Berera}},\ }\href@noop {} {\bibfield  {journal} {\bibinfo  {journal} {Int.
  J. Mod. Phys. A}\ }\textbf {\bibinfo {volume} {24}},\ \bibinfo {pages} {2207}
  (\bibinfo {year} {2009})}\BibitemShut {NoStop}%
\bibitem [{\citenamefont {Berera}\ \emph {et~al.}(1998)\citenamefont {Berera},
  \citenamefont {Gleiser},\ and\ \citenamefont {Ramos}}]{Berera1998}%
  \BibitemOpen
  \bibfield  {author} {\bibinfo {author} {\bibfnamefont {A.}~\bibnamefont
  {Berera}}, \bibinfo {author} {\bibfnamefont {M.}~\bibnamefont {Gleiser}}, \
  and\ \bibinfo {author} {\bibfnamefont {R.~O.}\ \bibnamefont {Ramos}},\
  }\href@noop {} {\bibfield  {journal} {\bibinfo  {journal} {Phys. Rev. D}\
  }\textbf {\bibinfo {volume} {58}},\ \bibinfo {pages} {123508} (\bibinfo
  {year} {1998})}\BibitemShut {NoStop}%
\bibitem [{\citenamefont {Yokoyama}\ and\ \citenamefont
  {Linde}(1999)}]{Yokoyama1999}%
  \BibitemOpen
  \bibfield  {author} {\bibinfo {author} {\bibfnamefont {J.}~\bibnamefont
  {Yokoyama}}\ and\ \bibinfo {author} {\bibfnamefont {A.}~\bibnamefont
  {Linde}},\ }\href@noop {} {\bibfield  {journal} {\bibinfo  {journal} {Phys.
  Rev. D}\ }\textbf {\bibinfo {volume} {60}},\ \bibinfo {pages} {083509}
  (\bibinfo {year} {1999})}\BibitemShut {NoStop}%
\bibitem [{\citenamefont {Bastero-Gil}\ \emph {et~al.}(2011)\citenamefont
  {Bastero-Gil}, \citenamefont {Berera},\ and\ \citenamefont
  {Rosa}}]{Bastero-Gil2011}%
  \BibitemOpen
  \bibfield  {author} {\bibinfo {author} {\bibfnamefont {M.}~\bibnamefont
  {Bastero-Gil}}, \bibinfo {author} {\bibfnamefont {A.}~\bibnamefont {Berera}},
  \ and\ \bibinfo {author} {\bibfnamefont {J.~G.}\ \bibnamefont {Rosa}},\
  }\href@noop {} {\bibfield  {journal} {\bibinfo  {journal} {Phys. Rev. D}\
  }\textbf {\bibinfo {volume} {84}},\ \bibinfo {pages} {103503} (\bibinfo
  {year} {2011})}\BibitemShut {NoStop}%
\bibitem [{\citenamefont {Arkani-Hamed}\ \emph {et~al.}(2001)\citenamefont
  {Arkani-Hamed}, \citenamefont {Cohen},\ and\ \citenamefont
  {Georgi}}]{Arkani-Hamed2001}%
  \BibitemOpen
  \bibfield  {author} {\bibinfo {author} {\bibfnamefont {N.}~\bibnamefont
  {Arkani-Hamed}}, \bibinfo {author} {\bibfnamefont {A.~G.}\ \bibnamefont
  {Cohen}}, \ and\ \bibinfo {author} {\bibfnamefont {H.}~\bibnamefont
  {Georgi}},\ }\href@noop {} {\bibfield  {journal} {\bibinfo  {journal} {Phys.
  Lett. B}\ }\textbf {\bibinfo {volume} {513}},\ \bibinfo {pages} {232}
  (\bibinfo {year} {2001})}\BibitemShut {NoStop}%
\bibitem [{\citenamefont {Schmaltz}\ and\ \citenamefont
  {Tucker-Smith}(2005)}]{Schmaltz2005}%
  \BibitemOpen
  \bibfield  {author} {\bibinfo {author} {\bibfnamefont {M.}~\bibnamefont
  {Schmaltz}}\ and\ \bibinfo {author} {\bibfnamefont {D.}~\bibnamefont
  {Tucker-Smith}},\ }\href@noop {} {\bibfield  {journal} {\bibinfo  {journal}
  {Annu. Rev. Nucl. Part. Sci.}\ }\textbf {\bibinfo {volume} {55}},\ \bibinfo
  {pages} {229} (\bibinfo {year} {2005})}\BibitemShut {NoStop}%
\bibitem [{\citenamefont {Kaplan}\ and\ \citenamefont
  {Weiner}(2004)}]{Kaplan2004}%
  \BibitemOpen
  \bibfield  {author} {\bibinfo {author} {\bibfnamefont {D.~E.}\ \bibnamefont
  {Kaplan}}\ and\ \bibinfo {author} {\bibfnamefont {N.}~\bibnamefont
  {Weiner}},\ }\href@noop {} {\bibfield  {journal} {\bibinfo  {journal} {J.
  Cosmol. Astropart. Phys.}\ }\textbf {\bibinfo {volume} {02}},\ \bibinfo
  {pages} {005} (\bibinfo {year} {2004})}\BibitemShut {NoStop}%
\bibitem [{\citenamefont {Arkani-Hamed}\ \emph {et~al.}(2003)\citenamefont
  {Arkani-Hamed}, \citenamefont {Cheng}, \citenamefont {Creminelli},\ and\
  \citenamefont {Randall}}]{Arkani-Hamed2003}%
  \BibitemOpen
  \bibfield  {author} {\bibinfo {author} {\bibfnamefont {N.}~\bibnamefont
  {Arkani-Hamed}}, \bibinfo {author} {\bibfnamefont {H.-C.}\ \bibnamefont
  {Cheng}}, \bibinfo {author} {\bibfnamefont {P.}~\bibnamefont {Creminelli}}, \
  and\ \bibinfo {author} {\bibfnamefont {L.}~\bibnamefont {Randall}},\
  }\href@noop {} {\bibfield  {journal} {\bibinfo  {journal} {J. Cosmol.
  Astropart. Phys.}\ }\textbf {\bibinfo {volume} {07}},\ \bibinfo {pages} {003}
  (\bibinfo {year} {2003})}\BibitemShut {NoStop}%
\bibitem [{\citenamefont {Peiris}\ \emph {et~al.}(2007)\citenamefont {Peiris},
  \citenamefont {Baumann}, \citenamefont {Friedman},\ and\ \citenamefont
  {Cooray}}]{Peiris2007}%
  \BibitemOpen
  \bibfield  {author} {\bibinfo {author} {\bibfnamefont {H.}~\bibnamefont
  {Peiris}}, \bibinfo {author} {\bibfnamefont {D.}~\bibnamefont {Baumann}},
  \bibinfo {author} {\bibfnamefont {B.}~\bibnamefont {Friedman}}, \ and\
  \bibinfo {author} {\bibfnamefont {A.}~\bibnamefont {Cooray}},\ }\href@noop {}
  {\bibfield  {journal} {\bibinfo  {journal} {Phys. Rev. D}\ }\textbf {\bibinfo
  {volume} {76}},\ \bibinfo {pages} {103517} (\bibinfo {year}
  {2007})}\BibitemShut {NoStop}%
\bibitem [{\citenamefont
  {Spali{\'n}ski}(2007{\natexlab{a}})}]{Spalinski2007-3}%
  \BibitemOpen
  \bibfield  {author} {\bibinfo {author} {\bibfnamefont {M.}~\bibnamefont
  {Spali{\'n}ski}},\ }\href@noop {} {\bibfield  {journal} {\bibinfo  {journal}
  {Phys. Lett. B}\ }\textbf {\bibinfo {volume} {650}},\ \bibinfo {pages} {313}
  (\bibinfo {year} {2007}{\natexlab{a}})}\BibitemShut {NoStop}%
\bibitem [{\citenamefont {Bessada}\ \emph {et~al.}(2009)\citenamefont
  {Bessada}, \citenamefont {Kinney},\ and\ \citenamefont
  {Tzirakis}}]{Bessada2009}%
  \BibitemOpen
  \bibfield  {author} {\bibinfo {author} {\bibfnamefont {D.}~\bibnamefont
  {Bessada}}, \bibinfo {author} {\bibfnamefont {W.~H.}\ \bibnamefont {Kinney}},
  \ and\ \bibinfo {author} {\bibfnamefont {K.}~\bibnamefont {Tzirakis}},\
  }\href@noop {} {\bibfield  {journal} {\bibinfo  {journal} {J. Cosmol.
  Astropart. Phys.}\ }\textbf {\bibinfo {volume} {09}},\ \bibinfo {pages} {031}
  (\bibinfo {year} {2009})}\BibitemShut {NoStop}%
\bibitem [{\citenamefont {Muslimov}(1990)}]{Muslimov1990}%
  \BibitemOpen
  \bibfield  {author} {\bibinfo {author} {\bibfnamefont {A.~G.}\ \bibnamefont
  {Muslimov}},\ }\href@noop {} {\bibfield  {journal} {\bibinfo  {journal}
  {Class. Quant. Gravity}\ }\textbf {\bibinfo {volume} {7}},\ \bibinfo {pages}
  {231} (\bibinfo {year} {1990})}\BibitemShut {NoStop}%
\bibitem [{\citenamefont {Salopek}\ and\ \citenamefont
  {Bond}(1990)}]{Salopek1990}%
  \BibitemOpen
  \bibfield  {author} {\bibinfo {author} {\bibfnamefont {D.~S.}\ \bibnamefont
  {Salopek}}\ and\ \bibinfo {author} {\bibfnamefont {J.~R.}\ \bibnamefont
  {Bond}},\ }\href@noop {} {\bibfield  {journal} {\bibinfo  {journal} {Phys.
  Rev. D}\ }\textbf {\bibinfo {volume} {42}},\ \bibinfo {pages} {3936}
  (\bibinfo {year} {1990})}\BibitemShut {NoStop}%
\bibitem [{\citenamefont {Kinney}(1997)}]{Kinney1997}%
  \BibitemOpen
  \bibfield  {author} {\bibinfo {author} {\bibfnamefont {W.~H.}\ \bibnamefont
  {Kinney}},\ }\href@noop {} {\bibfield  {journal} {\bibinfo  {journal} {Phys.
  Rev. D}\ }\textbf {\bibinfo {volume} {56}},\ \bibinfo {pages} {2002}
  (\bibinfo {year} {1997})}\BibitemShut {NoStop}%
\bibitem [{\citenamefont {Spali{\'n}ski}(2007{\natexlab{b}})}]{Spalinski2007}%
  \BibitemOpen
  \bibfield  {author} {\bibinfo {author} {\bibfnamefont {M.}~\bibnamefont
  {Spali{\'n}ski}},\ }\href@noop {} {\bibfield  {journal} {\bibinfo  {journal}
  {J. Cosmol. Astropart. Phys.}\ }\textbf {\bibinfo {volume} {04}},\ \bibinfo
  {pages} {018} (\bibinfo {year} {2007}{\natexlab{b}})}\BibitemShut {NoStop}%
\bibitem [{\citenamefont {Miranda}\ \emph {et~al.}(2012)\citenamefont
  {Miranda}, \citenamefont {Hu},\ and\ \citenamefont {Adshead}}]{Miranda2012}%
  \BibitemOpen
  \bibfield  {author} {\bibinfo {author} {\bibfnamefont {V.}~\bibnamefont
  {Miranda}}, \bibinfo {author} {\bibfnamefont {W.}~\bibnamefont {Hu}}, \ and\
  \bibinfo {author} {\bibfnamefont {P.}~\bibnamefont {Adshead}},\ }\href@noop
  {} {\bibfield  {journal} {\bibinfo  {journal} {Phys. Rev. D}\ }\textbf
  {\bibinfo {volume} {86}},\ \bibinfo {pages} {063529} (\bibinfo {year}
  {2012})}\BibitemShut {NoStop}%
\bibitem [{\citenamefont {Amani}\ \emph {et~al.}(2017)\citenamefont {Amani},
  \citenamefont {Rezazadeh}, \citenamefont {Abdolmaleki},\ and\ \citenamefont
  {Karami}}]{Amani2017}%
  \BibitemOpen
  \bibfield  {author} {\bibinfo {author} {\bibfnamefont {R.}~\bibnamefont
  {Amani}}, \bibinfo {author} {\bibfnamefont {K.}~\bibnamefont {Rezazadeh}},
  \bibinfo {author} {\bibfnamefont {A.}~\bibnamefont {Abdolmaleki}}, \ and\
  \bibinfo {author} {\bibfnamefont {K.}~\bibnamefont {Karami}},\ }\href@noop {}
  {\bibfield  {journal} {\bibinfo  {journal} {In preparation}\ } (\bibinfo
  {year} {2017})}\BibitemShut {NoStop}%
\bibitem [{\citenamefont {Liddle}\ and\ \citenamefont
  {Leach}(2003)}]{Liddle2003}%
  \BibitemOpen
  \bibfield  {author} {\bibinfo {author} {\bibfnamefont {A.~R.}\ \bibnamefont
  {Liddle}}\ and\ \bibinfo {author} {\bibfnamefont {S.~M.}\ \bibnamefont
  {Leach}},\ }\href@noop {} {\bibfield  {journal} {\bibinfo  {journal} {Phys.
  Rev. D}\ }\textbf {\bibinfo {volume} {68}},\ \bibinfo {pages} {103503}
  (\bibinfo {year} {2003})}\BibitemShut {NoStop}%
\bibitem [{\citenamefont {Dodelson}\ and\ \citenamefont
  {Hui}(2003)}]{Dodelson2003}%
  \BibitemOpen
  \bibfield  {author} {\bibinfo {author} {\bibfnamefont {S.}~\bibnamefont
  {Dodelson}}\ and\ \bibinfo {author} {\bibfnamefont {L.}~\bibnamefont {Hui}},\
  }\href@noop {} {\bibfield  {journal} {\bibinfo  {journal} {Phys. Rev. Lett.}\
  }\textbf {\bibinfo {volume} {91}},\ \bibinfo {pages} {131301} (\bibinfo
  {year} {2003})}\BibitemShut {NoStop}%
\bibitem [{\citenamefont {Ford}(1987)}]{Ford1987}%
  \BibitemOpen
  \bibfield  {author} {\bibinfo {author} {\bibfnamefont {L.~H.}\ \bibnamefont
  {Ford}},\ }\href@noop {} {\bibfield  {journal} {\bibinfo  {journal} {Phys.
  Rev. D}\ }\textbf {\bibinfo {volume} {35}},\ \bibinfo {pages} {2955}
  (\bibinfo {year} {1987})}\BibitemShut {NoStop}%
\bibitem [{\citenamefont {Aharony}\ \emph {et~al.}(2000)\citenamefont
  {Aharony}, \citenamefont {Gubser}, \citenamefont {Maldacena}, \citenamefont
  {Ooguri},\ and\ \citenamefont {Oz}}]{Aharony2000}%
  \BibitemOpen
  \bibfield  {author} {\bibinfo {author} {\bibfnamefont {O.}~\bibnamefont
  {Aharony}}, \bibinfo {author} {\bibfnamefont {S.~S.}\ \bibnamefont {Gubser}},
  \bibinfo {author} {\bibfnamefont {J.}~\bibnamefont {Maldacena}}, \bibinfo
  {author} {\bibfnamefont {H.}~\bibnamefont {Ooguri}}, \ and\ \bibinfo {author}
  {\bibfnamefont {Y.}~\bibnamefont {Oz}},\ }\href@noop {} {\bibfield  {journal}
  {\bibinfo  {journal} {Phys. Rep.}\ }\textbf {\bibinfo {volume} {323}},\
  \bibinfo {pages} {183} (\bibinfo {year} {2000})}\BibitemShut {NoStop}%
\bibitem [{\citenamefont {Okada}\ \emph {et~al.}(2016)\citenamefont {Okada},
  \citenamefont {Senoguz},\ and\ \citenamefont {Shafi}}]{Okada2016}%
  \BibitemOpen
  \bibfield  {author} {\bibinfo {author} {\bibfnamefont {N.}~\bibnamefont
  {Okada}}, \bibinfo {author} {\bibfnamefont {V.~N.}\ \bibnamefont {Senoguz}},
  \ and\ \bibinfo {author} {\bibfnamefont {Q.}~\bibnamefont {Shafi}},\
  }\href@noop {} {\bibfield  {journal} {\bibinfo  {journal} {Turkish J. Phys.}\
  }\textbf {\bibinfo {volume} {40}},\ \bibinfo {pages} {150} (\bibinfo {year}
  {2016})}\BibitemShut {NoStop}%
\bibitem [{\citenamefont {Berera}\ and\ \citenamefont
  {Ramos}(2001)}]{Berera2001}%
  \BibitemOpen
  \bibfield  {author} {\bibinfo {author} {\bibfnamefont {A.}~\bibnamefont
  {Berera}}\ and\ \bibinfo {author} {\bibfnamefont {R.~O.}\ \bibnamefont
  {Ramos}},\ }\href@noop {} {\bibfield  {journal} {\bibinfo  {journal} {Phys.
  Rev. D}\ }\textbf {\bibinfo {volume} {63}},\ \bibinfo {pages} {103509}
  (\bibinfo {year} {2001})}\BibitemShut {NoStop}%
\bibitem [{\citenamefont {Berera}\ and\ \citenamefont
  {Ramos}(2005)}]{Berera2005}%
  \BibitemOpen
  \bibfield  {author} {\bibinfo {author} {\bibfnamefont {A.}~\bibnamefont
  {Berera}}\ and\ \bibinfo {author} {\bibfnamefont {R.~O.}\ \bibnamefont
  {Ramos}},\ }\href@noop {} {\bibfield  {journal} {\bibinfo  {journal} {Phys.
  Lett. B}\ }\textbf {\bibinfo {volume} {607}},\ \bibinfo {pages} {1} (\bibinfo
  {year} {2005})}\BibitemShut {NoStop}%
\bibitem [{\citenamefont {Cai}\ \emph {et~al.}(2011)\citenamefont {Cai},
  \citenamefont {Dent},\ and\ \citenamefont {Easson}}]{Cai2011}%
  \BibitemOpen
  \bibfield  {author} {\bibinfo {author} {\bibfnamefont {Y.-F.}\ \bibnamefont
  {Cai}}, \bibinfo {author} {\bibfnamefont {J.~B.}\ \bibnamefont {Dent}}, \
  and\ \bibinfo {author} {\bibfnamefont {D.~A.}\ \bibnamefont {Easson}},\
  }\href@noop {} {\bibfield  {journal} {\bibinfo  {journal} {Phys. Rev. D}\
  }\textbf {\bibinfo {volume} {83}},\ \bibinfo {pages} {101301} (\bibinfo
  {year} {2011})}\BibitemShut {NoStop}%
\bibitem [{\citenamefont {Gleiser}\ and\ \citenamefont
  {Ramos}(1994)}]{Gleiser1994}%
  \BibitemOpen
  \bibfield  {author} {\bibinfo {author} {\bibfnamefont {M.}~\bibnamefont
  {Gleiser}}\ and\ \bibinfo {author} {\bibfnamefont {R.~O.}\ \bibnamefont
  {Ramos}},\ }\href@noop {} {\bibfield  {journal} {\bibinfo  {journal} {Phys.
  Rev. D}\ }\textbf {\bibinfo {volume} {50}},\ \bibinfo {pages} {2441}
  (\bibinfo {year} {1994})}\BibitemShut {NoStop}%
\bibitem [{\citenamefont {Zhang}\ and\ \citenamefont {Zhu}(2014)}]{Zhang2014}%
  \BibitemOpen
  \bibfield  {author} {\bibinfo {author} {\bibfnamefont {X.-M.}\ \bibnamefont
  {Zhang}}\ and\ \bibinfo {author} {\bibfnamefont {J.-Y.}\ \bibnamefont
  {Zhu}},\ }\href@noop {} {\bibfield  {journal} {\bibinfo  {journal} {Phys.
  Rev. D}\ }\textbf {\bibinfo {volume} {90}},\ \bibinfo {pages} {123519}
  (\bibinfo {year} {2014})}\BibitemShut {NoStop}%
\bibitem [{\citenamefont {Zhang}\ and\ \citenamefont {Zhu}(2015)}]{Zhang2015}%
  \BibitemOpen
  \bibfield  {author} {\bibinfo {author} {\bibfnamefont {X.-M.}\ \bibnamefont
  {Zhang}}\ and\ \bibinfo {author} {\bibfnamefont {J.-Y.}\ \bibnamefont
  {Zhu}},\ }\href@noop {} {\bibfield  {journal} {\bibinfo  {journal} {Phys.
  Rev. D}\ }\textbf {\bibinfo {volume} {91}},\ \bibinfo {pages} {063510}
  (\bibinfo {year} {2015})}\BibitemShut {NoStop}%
\bibitem [{\citenamefont {Spali{\'n}ski}(2008)}]{Spalinski2008}%
  \BibitemOpen
  \bibfield  {author} {\bibinfo {author} {\bibfnamefont {M.}~\bibnamefont
  {Spali{\'n}ski}},\ }\href@noop {} {\bibfield  {journal} {\bibinfo  {journal}
  {J. Cosmol. Astropart. Phys.}\ }\textbf {\bibinfo {volume} {04}},\ \bibinfo
  {pages} {002} (\bibinfo {year} {2008})}\BibitemShut {NoStop}%
\bibitem [{\citenamefont {Tsujikawa}\ \emph {et~al.}(2013)\citenamefont
  {Tsujikawa}, \citenamefont {Ohashi}, \citenamefont {Kuroyanagi},\ and\
  \citenamefont {{De Felice}}}]{Tsujikawa2013}%
  \BibitemOpen
  \bibfield  {author} {\bibinfo {author} {\bibfnamefont {S.}~\bibnamefont
  {Tsujikawa}}, \bibinfo {author} {\bibfnamefont {J.}~\bibnamefont {Ohashi}},
  \bibinfo {author} {\bibfnamefont {S.}~\bibnamefont {Kuroyanagi}}, \ and\
  \bibinfo {author} {\bibfnamefont {A.}~\bibnamefont {{De Felice}}},\
  }\href@noop {} {\bibfield  {journal} {\bibinfo  {journal} {Phys. Rev. D}\
  }\textbf {\bibinfo {volume} {88}},\ \bibinfo {pages} {023529} (\bibinfo
  {year} {2013})}\BibitemShut {NoStop}%
\bibitem [{\citenamefont {Moss}\ and\ \citenamefont {Xiong}(2006)}]{Moss2006}%
  \BibitemOpen
  \bibfield  {author} {\bibinfo {author} {\bibfnamefont {I.~G.}\ \bibnamefont
  {Moss}}\ and\ \bibinfo {author} {\bibfnamefont {C.}~\bibnamefont {Xiong}},\
  }\href@noop {} {\bibfield  {journal} {\bibinfo  {journal} {arXiv preprint
  hep-ph/0603266}\ } (\bibinfo {year} {2006})}\BibitemShut {NoStop}%
\bibitem [{\citenamefont {Morikawa}\ and\ \citenamefont
  {Sasaki}(1984)}]{Morikawa1984}%
  \BibitemOpen
  \bibfield  {author} {\bibinfo {author} {\bibfnamefont {M.}~\bibnamefont
  {Morikawa}}\ and\ \bibinfo {author} {\bibfnamefont {M.}~\bibnamefont
  {Sasaki}},\ }\href@noop {} {\bibfield  {journal} {\bibinfo  {journal}
  {Progress of Theoretical Physics}\ }\textbf {\bibinfo {volume} {72}},\
  \bibinfo {pages} {782} (\bibinfo {year} {1984})}\BibitemShut {NoStop}%
\bibitem [{\citenamefont {Berera}\ and\ \citenamefont
  {Ramos}(2003)}]{Berera2003}%
  \BibitemOpen
  \bibfield  {author} {\bibinfo {author} {\bibfnamefont {A.}~\bibnamefont
  {Berera}}\ and\ \bibinfo {author} {\bibfnamefont {R.~O.}\ \bibnamefont
  {Ramos}},\ }\href@noop {} {\bibfield  {journal} {\bibinfo  {journal} {Physics
  Letters B}\ }\textbf {\bibinfo {volume} {567}},\ \bibinfo {pages} {294}
  (\bibinfo {year} {2003})}\BibitemShut {NoStop}%
\bibitem [{\citenamefont {Zhang}(2009)}]{Zhang2009}%
  \BibitemOpen
  \bibfield  {author} {\bibinfo {author} {\bibfnamefont {Y.}~\bibnamefont
  {Zhang}},\ }\href@noop {} {\bibfield  {journal} {\bibinfo  {journal} {Journal
  of Cosmology and Astroparticle Physics}\ }\textbf {\bibinfo {volume}
  {2009}},\ \bibinfo {pages} {023} (\bibinfo {year} {2009})}\BibitemShut
  {NoStop}%
\bibitem [{\citenamefont {Hall}\ and\ \citenamefont {Moss}(2005)}]{Hall2005}%
  \BibitemOpen
  \bibfield  {author} {\bibinfo {author} {\bibfnamefont {L.~M.}\ \bibnamefont
  {Hall}}\ and\ \bibinfo {author} {\bibfnamefont {I.~G.}\ \bibnamefont
  {Moss}},\ }\href@noop {} {\bibfield  {journal} {\bibinfo  {journal} {Physical
  Review D}\ }\textbf {\bibinfo {volume} {71}},\ \bibinfo {pages} {023514}
  (\bibinfo {year} {2005})}\BibitemShut {NoStop}%
\bibitem [{\citenamefont {Hall}\ and\ \citenamefont {Peiris}(2008)}]{Hall2008}%
  \BibitemOpen
  \bibfield  {author} {\bibinfo {author} {\bibfnamefont {L.~M.}\ \bibnamefont
  {Hall}}\ and\ \bibinfo {author} {\bibfnamefont {H.~V.}\ \bibnamefont
  {Peiris}},\ }\href@noop {} {\bibfield  {journal} {\bibinfo  {journal}
  {Journal of Cosmology and Astroparticle Physics}\ }\textbf {\bibinfo {volume}
  {2008}},\ \bibinfo {pages} {027} (\bibinfo {year} {2008})}\BibitemShut
  {NoStop}%
\bibitem [{\citenamefont {Motaharfar}\ \emph {et~al.}(2018)\citenamefont
  {Motaharfar}, \citenamefont {Kamali},\ and\ \citenamefont
  {Ramos}}]{Motaharfar2018}%
  \BibitemOpen
  \bibfield  {author} {\bibinfo {author} {\bibfnamefont {M.}~\bibnamefont
  {Motaharfar}}, \bibinfo {author} {\bibfnamefont {V.}~\bibnamefont {Kamali}},
  \ and\ \bibinfo {author} {\bibfnamefont {R.~O.}\ \bibnamefont {Ramos}},\
  }\href@noop {} {\bibfield  {journal} {\bibinfo  {journal} {arXiv preprint
  arXiv:1810.02816}\ } (\bibinfo {year} {2018})}\BibitemShut {NoStop}%
\bibitem [{\citenamefont {Das}(2018)}]{Das2018}%
  \BibitemOpen
  \bibfield  {author} {\bibinfo {author} {\bibfnamefont {S.}~\bibnamefont
  {Das}},\ }\href@noop {} {\bibfield  {journal} {\bibinfo  {journal} {arXiv
  preprint arXiv:1810.05038}\ } (\bibinfo {year} {2018})}\BibitemShut {NoStop}%
\end{thebibliography}


%


\end{document}